\documentclass{aa}  
\usepackage{esvect} 
\usepackage{amsmath}
\usepackage{color}
\usepackage{graphicx}
\usepackage{txfonts}
\usepackage{hyperref} 
\usepackage{comment} 
\usepackage{xcolor} 
\usepackage{hyperref} 
\usepackage{tabularx}
\usepackage[flushleft]{threeparttable} 
\usepackage{makecell} 
\usepackage{dsfont} 
\usepackage{mathrsfs} 
\usepackage{esvect} 
\usepackage{amsmath} 
\usepackage{dsfont} 

\begin{document}

\title{Bayesian polarization calibration and imaging in very long baseline interferometry}

\titlerunning{Bayesian polarization calibration and imaging in VLBI}
\authorrunning{Kim et. al.}


\newcommand{\mj}[1]{\textcolor{red}{[MJ: #1]}}

\author{Jong-Seo Kim \inst{1}\fnmsep\thanks{jongkim@mpifr-bonn.mpg.de}
\and Jakob Roth \inst{2,3,4}
\and Jongho Park \inst{5}
\and Jack D. Livingston \inst{1}
\and Philipp Arras \inst{3}
\and Torsten A. En{\ss}lin \inst{3,6}
\and Michael Janssen \inst{1,7}
\and J. Anton Zensus \inst{1}
\and Andrei P. Lobanov \inst{1,8}
}

\institute{    
	    Max-Planck-Institut f\"ur Radioastronomie, Auf dem H\"ugel 69, D-53121 Bonn, Germany
        \and
        Max Planck Computing and Data Facility, Gießenbachstr. 2, 85748 Garching, Germany
        \and
        Max-Planck-Institut f\"ur Astrophysik, Karl-Schwarzschild-Str. 1, 85748 Garching, Germany
        \and
	    Technische Universität M\"unchen (TUM), Boltzmannstr. 3, 85748 Garching, Germany
        \and
        School of Space Research, Kyung Hee University, 1732, Deogyeong-daero, Giheung-gu, Yongin-si, Gyeonggi-do 17104, Republic of Korea
        \and
        Ludwig-Maximilians-Universit\"at, Geschwister-Scholl-Platz 1, 80539 Munich, Germany     
        \and
        Department of Astrophysics, Institute for Mathematics, Astrophysics and Particle Physics (IMAPP), Radboud University, P.O. Box 9010, 6500 GL Nijmegen, The Netherlands
        \and 
        Institut für Experimentalphysik, Universit\"at Hamburg, Luruper Chaussee 149, 22761, Hamburg, Germany 
         }
   \date{Received 20 November 2025 / accepted 25 May 2026}

 
  \abstract
   {Extracting polarimetric information from very long baseline interferometry (VLBI) data is demanding but vital to understanding the synchrotron radiation process and the magnetic fields of celestial objects, such as active galactic nuclei (AGNs). However, conventional \texttt{CLEAN}-based calibration and imaging methods provide suboptimal resolution without uncertainty estimation of calibration solutions, while requiring manual steering from an experienced user.}
   {We present a Bayesian polarization calibration and imaging method for millimeter and centimeter VLBI datasets, which explores the posterior distribution of antenna-based gains, polarization leakages, and polarimetric images jointly from precalibrated data. To validate our method, we compare our results with the \texttt{CLEAN} and regularized maximum likelihood (RML)-based software \texttt{ehtim}.}
   {We reconstructed the posterior distribution of Stokes images, gains, and leakages from real and synthetic datasets in the framework of Bayesian imaging software \texttt{resolve} using variational inference methods. Polarization constraints are enforced in the model. Furthermore, polarization calibration with several sources and multiple intermediate frequencies (IFs) is supported in order to maximize the parallactic angle coverage and identify instrumental corruptions per IF.}
   {We demonstrate our calibration and imaging method with observations of the quasar 3C273 with the Very Long Baseline Array (VLBA) at 15 GHz and the blazar OJ287 with the Global Millimeter VLBI Array (GMVA) and the Atacama Large Millimeter/submillimeter Array (ALMA) at 86 GHz. Compared to the \texttt{CLEAN} method, our approach provides physically realistic images that satisfy the positivity of the total intensity and polarization constraints and can reconstruct complex source structures composed of various spatial scales. In contrast to conventional imaging and calibration methods, our method systematically accounts for calibration uncertainties in the final images and provides uncertainties of Stokes images and calibration solutions.}
   {Our Bayesian polarization calibration and imaging method explores the posterior distribution of calibration solutions and reconstructs physically plausible high-resolution images from VLBI data. The automated Bayesian approach for calibration and imaging will be able to obtain high-fidelity polarimetric images using high-quality data from next-generation radio arrays. The pipeline developed for this work is publicly available.}


   \maketitle
%

\section{Introduction}

Polarization provides valuable information about astronomical objects and the ambient medium in astronomy, as it is one of the most direct ways to access information about astrophysical magnetic fields. In conjunction with very long baseline interferometry (VLBI), which is able to achieve nominal resolutions of $\sim20\,\mu$as at 1 mm \citep{EHT_2019_M87_paper1} and $\sim15\,\mu$as at 345 GHz \citep{EHT_2024_345GHz}, we have the ability to probe extreme magneto-ionic environments. 

Polarimetric studies with VLBI have revealed the presence of helical and toroidal magnetic fields in the jet of active galactic nuclei (AGN) \citep{Asada_2008_3C273,Hovatta_2012,Gabuzda_2017}, the magnetic field of the supermassive black holes M87* and Sgr A* \citep{EHT_2021_M87_polarimetry, EHT_2024_SGRA_polarimetry, EHT_2025_M87_2021}, and an alignment between polarization and the jets of AGN, suggesting the presence of magnetizing shocks in AGN jets \citep{Lister_2005,Pushkarev_2023}. However, polarization calibration of VLBI data is challenging, as polarization signals typically have S/N that are an order of magnitude lower than for total intensity. 

Polarized data after precalibration can still have significant residual corruption from leakage between one polarization feed to another, known as polarization leakage or "D-terms". Moreover, the complex source structure of calibrators in VLBI, especially at millimeter-wavelengths, hinders the estimation of this polarization leakage. The conventional polarization calibration approach in radio interferometry is observing a static point-like calibrator with known polarization properties and identifying leakage corruptions by using the information about the calibrator and disentangling source polarization from the instrumentation contributions that are being solved by the time-varying polarization of the feed angle rotation. However, due to the high angular resolution, it is unusual to find static point-like calibrators for polarization calibration in VLBI \citep{Cotton_1993}.

\citet{Leppanen_1995_LPCAL} introduced the \texttt{LPCAL} software to infer polarization leakage corruptions in VLBI datasets. The \texttt{LPCAL} software utilizes a polarization prior model with a set of components that assume that the linear polarization of a component is proportional to the total intensity of the corresponding component, known as the similarity approximation \citep{Cotton_1993, Leppanen_1995_LPCAL}. The simple and effective prior is advantageous to determine leakage corruption from sparse and noisy VLBI datasets due to the small number of degrees of freedom in the model. 

However, the similarity approximation in \texttt{LPCAL} is not valid for sources with a complex polarization structure \citep{Cotton_1993}. Furthermore, multisource calibration is not directly supported \citep[see][for an example of a multisource method with \texttt{LPCAL}.]{Lister_2005} and D-terms are considered static in time and frequency for each intermediate frequency (IF) by default in \texttt{LPCAL}. Recently, the new \texttt{CLEAN}-based polarization calibration software packages \texttt{GPCAL} \citep{Park_2021_GPCAL} and \texttt{PolSolve} \citep{Martividal_2021_PolSolve} have been able to employ reconstructed Stokes Q and U images as priors in leakage calibration, a technique called polarization self-calibration. These software support multisource models to increase parallactic angle coverage, and are able to solve for D-terms that vary with time and frequency.

Nevertheless, \texttt{GPCAL} and \texttt{PolSolve} still rely on the conventional \texttt{CLEAN} deconvolution algorithm.
Although \texttt{CLEAN} is easy to use and converges robustly, it comes with several limitations.
It requires user-dependent inputs, such as CLEAN windows and weighting schemes, which may introduce biases into the final results.
Furthermore, \texttt{CLEAN} assumes that the sky is a collection of point sources (usually of the size of the synthesized beam) or Gaussian blobs in multiscale \texttt{CLEAN} \citep{Cornwell_2008}, but has no explicit notion of an extended structure. This assumption may lead to severe artifacts on small scales that are corrected by a convolution with a restoring beam. This convolution results in suboptimal resolution, and the potential of radio interferometric observations to super-resolve sources (in the high S/N regime) cannot be exploited. Additionally, the \texttt{CLEAN} restored image does not necessarily fit the data, but rather a tapered version of it. Consequently, the CLEAN image model for leakage calibration is not suitable for radio interferometric data with a complex source structure.
Moreover, conventional \texttt{CLEAN}-based self-calibration is performed iteratively by switching between flagging and manually choosing gain solution intervals, resulting in inconsistent calibration solutions \citep{Martividal_2008, Popkov_2021} and a lack of reproducibility. The conventional approach utilizes the sum of the \texttt{CLEAN} components without convolution as a model in self-calibration. This may imprint spurious small-scale structures in the self-calibrated data. Lastly, calibration uncertainties are not taken into account in the final results, and uncertainty estimation is not supported.

Modern forward modeling imaging algorithms provide us with a new approach to overcome the limitations of \texttt{CLEAN}-based polarization calibration methods \citep{2020_Birdi_polcal_SARA, 2021_Pesce_DMC, EHT_2021_M87_polarimetry, EHT_2024_SGRA_polarimetry}. Explicit regularizers and prior assumptions in forward modeling algorithms enable us to infer more robust leakage solutions from sparse VLBI datasets. For example, polarization constraints can be enforced in the polarization imaging model to avoid unphysical polarimetric images \citep{Birdi_2018_Polarized_SARA}. Recent instrumental advancement in VLBI accomplishes significantly improved S/N and more wideband observations. In addition, the latest algorithmic developments offer methods for solving highly degenerate inverse problems efficiently. A modern polarization calibration and imaging pipeline that utilizes the full potential of the data is highly desirable. 

In this work, we introduce a novel Bayesian calibration and imaging method using the Bayesian imaging software \texttt{resolve}. The posterior distribution of antenna-based gains, D-terms, and Stokes images are explored jointly using variational inference algorithms \citep{Knollmueller_2019_MGVI, Frank_2021_geoVI}. Recently, the polarization constraint $(I \geq \sqrt{Q^{2}+U^{2}+V^{2}})$ is encoded in the the \texttt{resolve} polarization imaging model \citep{Arras_25}. We utilize this polarization imaging model and incorporate polarization calibration and gain self-calibration directly into the imaging process. Furthermore, multisource and multi-IF polarization calibration are supported in a probabilistic framework.

This article is structured as follows. In \autoref{Sect:Method}, we explain the conventional \texttt{CLEAN}-based polarization calibration method and our Bayesian method. In \autoref{Sect:synthetic_data} and \autoref{Sect:real_data}, we validate the method with synthetic and real datasets, respectively. We summarize our results in \autoref{Sect:conclusion}.

\section{Method}\label{Sect:Method}
\subsection{Radio interferometer measurement equation (RIME)}

A radio interferometer measures Fourier components of the sky brightness distribution instead of imaging the sky directly. According to the van Cittert-Zernike theorem, the two-point correlation function of the signals recorded by two antennas, $i$ and $j$, called visibility data $V_{ij}$, is given by the Fourier-transformed sky brightness distribution $I$ for total intensity under the assumption that the field of view is small \citep{Hamaker_1996, Smirnov_2011, Thompson_2017} 
\begin{align} \label{eq:measurement_equation}
   V(u,v)_{ij} = \int_{-\infty}^{\infty}\int_{-\infty}^{\infty} I(x,y) \, e^{-2\pi \mathbf{i} (ux+vy)} dx \,dy = \mathbb{FT} [I],
\end{align}
where $(u,v)$ are the Fourier domain coordinates, $(x,y)$ are the image domain coordinates, and $\mathbb{FT}$ is the Fourier transform operator. \\

Since the observed data are incomplete and corrupted by atmospheric and instrumental effects, we obtain a low-fidelity image, also known as the dirty map, from a direct Fourier transform of the visibility data. Inferring the real source structure from radio interferometric data is an ill-posed inverse problem, and therefore a unique solution does not exist. Thus, additional assumptions, prior knowledge, or other regularizers of the solution space are required to obtain high-fidelity images from radio interferometric data in image reconstruction.

The measurement equation (Eq. \ref{eq:measurement_equation}) can be generalized for full polarization observation, and the instrumental and atmospheric data corruption can be described by Jones matrices \citep{Jones_1941}. 
As a result, the radio interferometer measurement equation (RIME) for the full polarimetric visibility matrix $\mathbf{V}$ on the basis of circular polarization is given by\citep{Smirnov_2011}

\begin{align} \label{eq:measurement_equation_with_gains}
   \mathbf{V}_{ij} = \mathbf{J}_i \, \biggl( \int_{-\infty}^{\infty}\int_{-\infty}^{\infty} \mathbf{I}(x,y) \, e^{-2\pi \mathbf{i} (ux+vy)} dx \,dy \biggr) \,  \mathbf{J}_j^{\dagger} + \mathbf{N}_{ij},
\end{align}
where $\mathbf{V}_{ij}$ is the visibility matrix consisting of four complex correlation functions by the signal from the right-hand circular polarization (RCP) $R$ and the signal from the left-hand circular polarization (LCP) $L$

\begin{equation}
    \mathbf{V}_{ij} = \begin{pmatrix} \,R_{i}R_{j}^{*} & R_{i}L_{j}^{*}\, \\ \,L_{i}R_{j}^{*} & L_{i}L_{j}^{*}\, \end{pmatrix},
\end{equation}

\noindent
where the asterisk $*$ denotes a complex conjugate,

\noindent
$\mathbf{J}_{i}$ is the Jones matrix that describes the data corruption of the antenna $i$ by instrumental and atmospheric effects  

\begin{equation}
    \mathbf{J}_{i} = \mathbf{G}_i \, \mathbf{D}_i \, \mathbf{P}_i = \begin{pmatrix} g^{R}_i & 0 \\ 0 & g^{L}_i \end{pmatrix} \begin{pmatrix} 1 & D^{R}_i \\ D^{L}_i & 1 \end{pmatrix} \begin{pmatrix} e^{-\mathbf{i} \phi_i} & 0 \\ 0 & e^{\mathbf{i} \phi_i} \end{pmatrix},
\end{equation}

\noindent
where $\mathbf{G}_{i}$ is the antenna-based gain matrix, $\mathbf{D}_{i}$ is the leakage (D-term) matrix describing the signal leakage between polarizers (e.g., $D^{R}_i$ is the signal leakage from LCP to RCP for antenna $i$),  $\mathbf{P}_{i}$ is the field rotation angle matrix, $\dagger$ denotes conjugate transposition, $\mathbf{N}_{ij}$ is the additive noise, and $\mathbf{I}$ is the polarimetric sky brightness distribution matrix, consisting of Stokes $I$, $Q$, $U$, and $V$

\begin{equation}
    \mathbf{I} =\begin{pmatrix} I+V & Q+\mathbf{i}U \\ Q-\mathbf{i}U & I-V \end{pmatrix}.
\end{equation}

The antenna field rotation angle $\phi$, representing the rotation of the receiver polarization feeds with respect to the source due to Earth's rotation, is defined as $\phi = f_{\text{el}} \theta_{\text{el}} + f_{\text{par}} \psi_{\text{par}} + \phi_{\text{off}}$, where $\theta_{\text{el}}$ is the elevation angle, $\psi_{\text{par}}$ is the parallactic angle, and $\phi_{\text{off}}$ is a constant offset. The field rotation angle depends on the antenna mount. Alt-azimuth (ALT-AZ) mounts with a Cassegrain focus have $f_{\text{par}} = 1$ and $f_{\text{el}} = 0$. The ALT-AZ mounts with a Nasmyth-Right-type focus have $f_{\text{par}} = 1$ and $f_{\text{el}} = 1$, and ALT-AZ mounts with a Nasmyth-Left-type focus have $f_{\text{par}} = 1$ and $f_{\text{el}} = -1$. More details regarding the antenna mount types can be found in Appendix C of \citet{Janssen_2019_rPICARD}.\\

We note that no real antenna feed responds to a single sense of perfectly circular polarization. The voltage induced in each nominally circularly polarized channel consists of two terms: the intended response of the feed and a smaller response to orthogonal polarization \citep{Roberts_1994}. As a result, the polarization state received by each feed is slightly elliptical rather than purely circular (e.g., \citealt{Thompson_2017}). The complex D-term coefficients parameterize this departure from ideal circular polarization: their amplitudes are related to the axial ratio of the polarization ellipse traced by each feed's effective response, and their phases describe the orientation of the ellipse on the sky \citep{Roberts_1994, Park_2023_GPCAL}.

At centimeter and shorter wavelengths, a typical circularly polarized feed consists of a horn followed by a polarizing device, such as a quarter-wave plate or a septum polarizer, that converts the intrinsically linearly polarized detector response to circular polarization \citep{Roberts_1994}. Imperfections in this conversion process—including differential amplitude and phase errors between the two orthogonal probes and deviations of the phase shift from the ideal $\pi$/2—give rise to the D-terms. Since the polarizer performance is inherently frequency-dependent, the resulting D-terms generally vary across the observing band, which motivates per-IF calibration for observations with a wide fractional bandwidth (\citealt{Park_2023_GPCAL}; see also Section 4.1). In addition, the D-terms are direction-dependent \citep{Smirnov_2011}, such that the effective leakage of an antenna can vary with time due to changes in the antenna pointing offset caused by winds and dish deformation \citep{Park_2023_GPCAL2}. \\

The polarization calibration and imaging model with the polarimetric visibility matrix data $\mathbf{V}$, Jones matrices $\mathbf{J_i}$, $\mathbf{J_j}$ for antennas, $i$ and $j$, and model visibilities ($\mathcal{RR}$, $\mathcal{RL}$, $\mathcal{LR}$, and $\mathcal{LL}$) from Stokes images is

\begin{align}\label{eq:polcal_imaging_model}
    R_{i}R_{j}^{*} &= g^{R}_i \, g^{R*}_{j} [e^{-\mathbf{i} \phi_i} \, \mathcal{RR} \, e^{\mathbf{i} \phi_j} + D^{R}_i \, e^{\mathbf{i} \phi_i} \, \mathcal{LR} \, e^{\mathbf{i}\phi_j} \notag \\
    &+ e^{-\mathbf{i}\phi_i} \, \mathcal{RL} \,e^{-\mathbf{i}\phi_j} \, D^{R*}_j + D^{R}_i e^{\mathbf{i}\phi_i} \, \mathcal{LL} e^{-\mathbf{i}\phi_j} \, D^{R*}_{j}] \notag \\
    R_{i}L_{j}^{*} &= g^{R}_i \, g^{L*}_{j} [e^{-\mathbf{i} \phi_i} \, \mathcal{RR} \, e^{\mathbf{i} \phi_j} D^{L*}_j + D^{R}_i \, e^{\mathbf{i} \phi_i} \, \mathcal{LR} \, e^{\mathbf{i}\phi_j} \, D^{L*}_j \notag \\
    &+ e^{-\mathbf{i}\phi_i} \, \mathcal{RL} \,e^{-\mathbf{i}\phi_j} + D^{R}_i e^{\mathbf{i}\phi_i} \, \mathcal{LL} e^{-\mathbf{i}\phi_j}] \notag \\
    L_{i}R_{j}^{*} &= g^{L}_i \, g^{R*}_{j} [D^{L}_i \, e^{-\mathbf{i} \phi_i} \, \mathcal{RR} \, e^{\mathbf{i} \phi_j} + e^{\mathbf{i} \phi_i} \, \mathcal{LR} \, e^{\mathbf{i}\phi_j} \notag \\
    &+ D^{L}_i \, e^{-\mathbf{i}\phi_i} \, \mathcal{RL} \,e^{-\mathbf{i}\phi_j} \, D^{R*}_j + e^{\mathbf{i}\phi_i} \, \mathcal{LL} e^{-\mathbf{i}\phi_j} \, D^{R*}_j] \notag\\
    L_{i}L_{j}^{*} &= g^{L}_i \, g^{L*}_{j} [D^{L}_i \, e^{-\mathbf{i} \phi_i} \, \mathcal{RR} \, e^{\mathbf{i} \phi_j} \, D^{L*}_j + e^{\mathbf{i} \phi_i} \, \mathcal{LR} \, e^{\mathbf{i}\phi_j} \, D^{L*}_j \notag \\
    &+ D^{L}_i \, e^{-\mathbf{i}\phi_i} \, \mathcal{RL} \,e^{-\mathbf{i}\phi_j} + e^{\mathbf{i}\phi_i} \, \mathcal{LL} e^{-\mathbf{i}\phi_j} ],
\end{align}

\noindent
where the model visibility matrix consists of Stokes images

\begin{equation}
     \mathbf{V}_{model} = \begin{pmatrix} \mathcal{RR} & \mathcal{RL} \\
     \mathcal{LR} & \mathcal{LL}
     \end{pmatrix} = \mathbb{FT} \left[
     \begin{pmatrix} I+V & Q+\mathbf{i}U \\ Q-\mathbf{i}U & I-V \end{pmatrix} \right].
\end{equation}

Polarization calibration and imaging are equivalent to estimating the gains $g$, D-terms $D$, Stokes images $I, Q, U,$ and $V$ from the visibility data $\mathbf{V}$. In the conventional \texttt{CLEAN}-based method, this polarimetric calibration and imaging problem in VLBI is solved in an iterative fashion since the problem is highly degenerate. In this work, we infer the gains $g$, leakages $D$, Stokes $I, Q, U,$ and $V$ jointly from precalibrated visibility matrix data $\mathbf{V}$ in a probabilistic approach.

\subsection{\texttt{CLEAN}-based polarization calibration method}

The \texttt{CLEAN}-based polarization calibration method assumes that the antenna gains have already been calibrated during the data preprocessing and imaging and/or self-calibration procedures. It also assumes that the antenna's field rotation angles have been corrected during the data preprocessing phase (see \citealt{Park_2021_GPCAL, Park_2023_GPCAL, Martividal_2021_PolSolve} for more details). With these assumptions, the cross-hand visibilities consist of both the source's intrinsic linear polarization terms and the terms associated with the antenna's polarimetric leakages and field rotation angles.

Determining leakages is not straightforward, as they need to be disentangled from the source polarization terms. One of the most widely used methods in the past assumes that the source's linear polarization emission is proportional to the total intensity structure within each "submodel", which consists of a group of neighboring total intensity CLEAN components. This method is known as the "similarity approximation" \citep{Cotton_1993, Leppanen_1995_LPCAL} and is implemented in \texttt{LPCAL}, a task within the Astronomical Image Processing System (\texttt{AIPS}; \cite{Greisen_2003_AIPS}).

While generally a good approximation, the similarity method faces challenges when dealing with very weakly polarized sources \citep[e.g.,][]{Park_2021_M87}, particularly when utilizing global VLBI observations at millimeter wavelengths, which provide ultra-high angular resolution. In such cases, the source's linear polarization structures are often complex, and the similarity approximation may not hold well \citep[e.g.,][]{EHT_2021_M87_polarimetry, EHT_2024_SGRA_polarimetry, Zhao_2022}.

To overcome this limitation, two methods have been developed: \texttt{GPCAL} \citep{Park_2021_GPCAL}, based on \texttt{AIPS} and \texttt{Difmap}, and \texttt{PolSolve} \citep{Martividal_2021_PolSolve}, based on \texttt{CASA} \citep{CASA_2022}. These methods conduct calibration as follows: 

\begin{enumerate}
    \item They derive the leakages using the similarity approximation, such as LPCAL, and remove them from the data.
    \item They perform imaging with \texttt{CLEAN} using the leakage-corrected data to obtain the source's Stokes $Q$ and $U$ CLEAN models.
    \item They use the original data to derive the leakage solutions again, this time using the source's linear polarization models obtained in the previous step, and remove the leakages using the updated solutions.
    \item They iteratively update the source's linear polarization models and leakage solutions by repeating steps 2 and 3 until the solutions converge.
\end{enumerate} 
\noindent These methods can also simultaneously utilize data from multiple calibrator sources, which typically leads to improved leakage calibration accuracy, as leakage solutions are not expected to vary between sources.

\subsection{Bayesian imaging software \texttt{resolve}}\label{sec:bayesimaging}

The open-source Bayesian imaging software \texttt{resolve}\footnote{\url{https://gitlab.mpcdf.mpg.de/ift/resolve}} treats radio interferometric imaging and calibration (Eq. \ref{eq:measurement_equation_with_gains}) as an inverse problem and computes a probabilistic solution for it. Thus, \texttt{resolve} computes the probability distribution of the sky brightness $\mathbf{I}$ given the measured visibilities. In \cite{Arras_2019}, a joint calibration and imaging approach was introduced, computing a posterior probability distribution for the sky brightness $\mathbf{I}$ and the antenna gains $\mathbf{G}$. \cite{Roth_23} extended this approach, incorporating direction-dependent effects in the antenna gains. A further extension to \texttt{resolve} was made in \cite{Arras_25}, enabling full Stokes imaging. In \cite{Roth_24}, the idea of major and minor cycles used in \texttt{CELAN}-based algorithms was adapted in a Bayesian version for the \texttt{resolve} framework.

Building on the full Stokes imaging capabilities, this work introduces polarization calibration and imaging to \texttt{resolve}. Thus, we compute the posterior distribution of the full Stokes sky brightness matrix $\mathbf{I}$ jointly with the posterior distributions of the antenna-based gain matrix $\mathbf{G}$, and the leakage matrix $\mathbf{D}$. The posterior distribution can be expressed via Bayes' theorem 
\begin{align} \label{eq:Bayes' thm}
\mathcal{P}(\mathbf{G}, \mathbf{D}, \mathbf{I}|\mathbf{V}) = \frac{\mathcal{P}(\mathbf{V}|\mathbf{G}, \mathbf{D}, \mathbf{I})\, \mathcal{P}(\mathbf{G}, \mathbf{D}, \mathbf{I})} {\mathcal{P}(\mathbf{V})},
\end{align}
in terms of the likelihood $\mathcal{P}(\mathbf{V}|\mathbf{G}, \mathbf{D}, \mathbf{I})$ and the prior $\mathcal{P}(\mathbf{G}, \mathbf{D}, \mathbf{I})$. In the following section \ref{sec:likelihood}, we discuss the likelihood model. In section \ref{sec:prior}, we outline the prior models, and in section \ref{sec:posterior}, we describe the algorithm for approximating the posterior distribution given the likelihood and prior.

\subsection{Likelihood distribution}\label{sec:likelihood}

As outlined in Section~\ref{sec:bayesimaging}, we approach the polarization calibration and imaging problem as a Bayesian inference task.
This entails defining the prior and likelihood to compute the posterior distribution, which represents the result of the Bayesian inference process.

Following the framework established by \citet{knoll_18}, we employ a coordinate transformation, or reparameterization trick (also known as inverse transform sampling), to introduce new parameters $\xi$.
This transformation ensures that the prior follows a standard normal distribution $\mathcal{P}(\xi) = \mathcal{G}(\xi, \mathds{1})$, integrating all prior knowledge into the likelihood component $\mathcal{P}(V | \xi)$.


The likelihood $\mathcal{P}(V | \xi)$ is composed of two components: $\mathcal{P}(\mathbf{V} | \mathbf{G}, \mathbf{D}, \mathbf{I})$ and a function of mapping $\xi$ to specific values of $\mathcal{G}$, $\mathcal{D}$, and $\mathcal{I}$.
The former component describes the data aspect of the likelihood, while the latter incorporates the prior information for the gains $g$, the leakages $D$, and the Stokes images $I, Q, U,$ and $V$. This section elaborates on the first component, while the following section will deal with the second.

Our data model integrates gain matrices $\mathbf{G}$, leakage matrices $\mathbf{D}$, field rotation angle matrices $\mathbf{P}$, and the polarization imaging model $\mathbf{I}$.
Upon setting $\mathbf{G}$, $\mathbf{D}$, $\mathbf{P}$, and $\mathbf{I}$, the visibility data $\mathbf{V}$ can be formulated as follows
\begin{align}
    \mathbf{V}_{ij} =  \mathbf{G}_i \, \mathbf{D}_i \, \mathbf{P}_i  \, \mathbb{FT} \, [\, \mathbf{I} \,] \,  \mathbf{P}_j^{\dagger} \, \mathbf{D}_j^{\dagger} \, \mathbf{G}_j^{\dagger}.
\end{align}

By encapsulating Jones matrices into the response function, the model can be simplified
\begin{align}
    \mathbf{V}_{ij} =  \mathbf{J}_i \, \mathbb{FT} \, [\, \mathbf{I} \,] \,  \mathbf{J}_j^{\dagger} = \mathbf{R}^{(g, D)} [\mathbf{I}] \, ,
\end{align}
where $\mathbf{R}^{(g, D)}$ is the response function consisting of the Fourier operator and Jones matrices.

It is important to note that field rotation angle matrices $\mathbf{P}$ are often precorrected during precalibration to stabilize the phase calibration in VLBI.\@
However, the field rotation angle matrix $\mathbf{P}$ and the D-term matrix $\mathbf{D}$ are not commutative.
Consequently, our model for precorrected field rotation angle data is
\begin{align}
    \mathbf{V}_{ij}^{\phi \, \text{cor.}} =  \mathbf{P}_i^{-1}  \mathbf{G}_i \, \mathbf{D}_i \, \mathbf{P}_i  \, \mathbb{FT} \, [\, \mathbf{I} \,] \,  \mathbf{P}_j^{\dagger} \, \mathbf{D}_j^{\dagger} \, \mathbf{G}_j^{\dagger} \, (\mathbf{P}_j^{\dagger})^{-1}.
\end{align}
This means we must reverse the precorrection of the field rotation angle in precalibrated data before adequately solving the Jones matrices, given the noncommutative properties of the matrices $\mathbf{G}$, $\mathbf{D}$, and $\mathbf{P}$.

Assuming additive Gaussian noise on the radio interferometric data, justified by the central limit theorem, the likelihood given the noise covariance $\mathbb{N}$ is expressed as
\begin{align}
    \mathcal{P}(\mathbf{V} | \mathbf{I}, \mathbf{G}, \mathbf{D}) = \mathcal{G} (\mathbf{V} - \mathbf{R}^{(g, D)} [\mathbf{I}] , \mathbb{N}).
\end{align}

In this work, the noise covariance is assumed to have a diagonal covariance; thus, the noise is uncorrelated. This likelihood formulation underscores how our method, unlike \texttt{CLEAN}-based polarization calibration methods, facilitates the simultaneous inference of calibration solutions and Stokes images.
Consequently, the uncertainty estimation of Stokes images inherently reflects uncertainties from the calibration, data noise, and incomplete UV-coverage.

\begin{table}[h]
    \noindent
    \caption{List of prior distributions in the Bayesian polarization calibration and imaging model.}
    \label{table:imaging_cal_prior}
	\begin{tabular}{llc}
		\multicolumn{3}{l}{\rule{8.7cm}{3\arrayrulewidth}} \\
        Model & Description & Prior \\
		\multicolumn{3}{l}{\rule{8.7cm}{2\arrayrulewidth}} \\
        \vspace{0.15cm}
		$s$ & Polarization imaging model & 2D $\mathcal{GP}(\xi_{s})$ \\ \vspace{0.15cm}
        $q$ & Polarization imaging model & 2D $\mathcal{GP}(\xi_{q})$ \\ \vspace{0.15cm}
        $u$ & Polarization imaging model & 2D $\mathcal{GP}(\xi_{u})$ \\ \vspace{0.15cm}
        $v$ & Polarization imaging model & 2D $\mathcal{GP}(\xi_{v})$ \\ \vspace{0.15cm}
        $\lambda$ & Lognormal amplitude gain & 1D $\mathcal{GP}(\xi_{\lambda})$ \\ \vspace{0.15cm}
        $\phi$ & Phase gain & 1D $\mathcal{GP}(\xi_{\phi})$ \\ \vspace{0.15cm}
         &  & or $\mathcal{N}(0, \sigma_{\phi}^{2})$ \\ \vspace{0.15cm}
        
        $a$ & Lognormal amplitude D-term & 
        $\mathcal{N}(m_{a}, \sigma_{a}^{2})$ \\ \vspace{0.15cm}
        $b$ & Phase D-term & $\mathcal{N}(0, \sigma_{b}^{2})$ \\
		\multicolumn{3}{l}{\rule{8.7cm}{3\arrayrulewidth}}
	\end{tabular}
\end{table}

\subsection{Polarization calibration and imaging prior model}\label{sec:prior}


For total intensity imaging, we utilize the \texttt{resolve} lognormal sky model, which builds on a Gaussian process in  \texttt{NIFTy} software\footnote{\url{https://gitlab.mpcdf.mpg.de/ift/nifty}}
\begin{equation}
    I(\vv{x}) = \text{exp}(s(\vv{x})), \, \, \,  s \curvearrowleft \mathscr{G}(s,S) \, ,
\end{equation}
where $S$ is the covariance matrix of the Gaussian process.\\

A priori, we assume homogeneous and isotropic statistics for the sky brightness. Due to the Wiener-Khinchin theorem \citep{Wiener_1949, Khinchin_1934}, this assumption allows us to represent the covariance matrix $S$ of the Gaussian process by a one-dimensional power spectrum $P_{s}$ in the Fourier domain.
As the degrees of freedom of the power spectrum scale linearly with the number of pixels in the image, this assumption makes it numerically affordable to infer spatial correlations even for high-dimensional $(N >10^6)$ radio images.
We transform the lognormal model so that the formal prior distribution is a standard normal distribution
\begin{equation} \label{eq:Gaussian_process}
    I(\vv{x}) = \text{exp}(s(\vv{x})) = \text{exp}(\mathbb{F}[\sqrt{P_{s}(\xi_s)}\xi_k]), \, \, \xi \curvearrowleft \mathcal{G}(\xi, \mathds{1}),
\end{equation}
with $P_{s}$ being our model for the power spectrum and $\xi_{s/k}$ the standard normally distributed parameters. As in previous \texttt{resolve} applications, we use a stochastic process-based nonparametric model for the power spectrum $P_{s}$. Importantly, this nonparametric model allows coverage of a wide range of possible correlation patterns, making it robustly applicable to a wide range of sources.

In this work, this Gaussian process prior is utilized for polarimetric imaging and gain models. A more detailed description of our generative Gaussian process model can be found in \autoref{Appendix_hyperparameter}.

\paragraph{Generative model for polarization imaging prior $\mathbf{I}$}
The generative model for the Stokes sky emission $\mathbf{I}$ closely follows the model presented in \citet{Arras_25}.
Regarded as a complex $2\times2$ matrix, $\mathbf{I}$ is defined by


\begin{align}
\label{eq:intro:X}
\mathbf I =\begin{pmatrix} \langle E_{i,r}E_{j,r}^{*}\rangle &\langle E_{i,r}E_{j,l}^{*}\rangle\\\langle E_{i,l}E_{j,r}^{*}\rangle &\langle E_{i,l}E_{j,l}^{*}\rangle \end{pmatrix} = \begin{pmatrix} I+V & Q+\mathbf{i} U\\Q-\mathbf{i} U & I-V \end{pmatrix},
\end{align}

\noindent
within the circular basis \citep{Smirnov_2011}.
Here, $E$ is the electric field and the indices $i,j$ denote antenna labels and $r,l$ refer to the respective circular feeds.
The matrix must meet specific constraints: (1) $\mathbf{I}$ is positive definite and Hermitian, (2) strictly positive total flux $I>0$, and (3) an upper bound on polarized emission $I^2 \geq Q^2+U^2+V^2$ \citep{Hamaker_1996,Smirnov_2011}.

\citet{Arras_25} identify the matrix exponential as a fitting parameterization for the polarized sky brightness distribution
\begin{align}
\mathbf I =e^{x} \coloneqq \exp \begin{pmatrix} s+v & q+\mathbf{i} u\\q-\mathbf{i} u & s-v \end{pmatrix},
\end{align}
where $s, q, u$, and $v$ are real numbers for each pixel and, in particular, can be both positive and negative. \\

Stokes $I,Q,U$, and $V$ can be obtained from 2D fields $s, q, u$, and $v$

\begin{align}\label{eq:pol_imaging_prior}
    I &= e^{s}\text{cosh} \,p, \hspace{2.2cm} Q = \frac{q}{p}e^{s}\text{sinh} \,p, \nonumber \\
    U &= \frac{u}{p}e^{s}\text{sinh} \,p, \hspace{2cm} V = \frac{v}{p} e^{s} \text{sinh} \,p,
\end{align}

\noindent
where $p \coloneqq \sqrt{q^{2}+u^{2}+v^{2}}$. \\

The total flux I is strictly positive in \autoref{eq:pol_imaging_prior} and $\mathbf{I}$ is Hermitian since the matrix exponential and the Hermitian conjugate commute. Furthermore, $\mathbf{I}$ is positive definite because the eigenvalues of a Hermitian matrix are real and the eigenvalues of the matrix exponential are the exponential of the eigenvalues. The determinant of $\mathbf{I}$ is the product of the eigenvalues, which is positive 
\begin{equation}
0 < \text{det} \, \mathbf{I} = I^{2}-Q^{2}-U^{2}-V^{2}.
\end{equation}

Thus, it proves that this parameterization satisfies all three conditions of $\mathbf{I}$.

To complete the generative model for $\mathbf{I}$, we need to define the models that generate the 2D fields $s, q, u$, and $v$.
We model each of those fields using a Gaussian process $(\mathcal{GP})$ with a nonparametric correlation kernel in \autoref{eq:Gaussian_process}, captured through generative models that convert standard-normal distributed latent parameters $\xi_s,\, \xi_q,\, \xi_u,\, \xi_v$ into the Gaussian process values $s(\xi_s),\, q(\xi_q),\, u(\xi_u),\, v(\xi_v)$ (following the methods in \citet{Arras_21}). In our polarization imaging prior model, the spatial correlation between pixels and the correlation between Stokes images are taken into account since the correlation kernels of $s, q, u$, and $v$ are inferred from the data. 

We note that our polarization imaging model is not well suited for data showing larger amplitudes in RL, LR visibilities than RR, LL visibilities. However, those cases are limited, and our model enforces the polarization constraint only in the image domain, not in the visibility domain.

\paragraph{Generative model for gain prior $g$}

The antenna-based gain prior consists of two Gaussian process models
\begin{align}
    g(t) = \text{exp}(\lambda(t)+\mathbf{i} \phi(t)),
\end{align}
where lognormal amplitude gain $\lambda$ and phase gain $\phi$ are 1D Gaussian process priors.

For the 1D gain prior, we use the same generative Gaussian process model in the imaging prior (see \autoref{eq:Gaussian_process}). Thus, the lognormal amplitude gain $\lambda(\xi_{\lambda})$ and phase gain $\phi(\xi_{\phi})$ can be described by the parameters of the standard-normal distributed model $\xi_{\lambda}$ and $\xi_{\phi}$. 

In radio interferometry the gain phase is inherently degenerate since we do not measure the absolute phase. To address the degeneracy, a standard approach is to choose a reference antenna with fixed gain phases in the polarization calibration. In our method, we can choose one reference antenna with high sensitivity or a short-baseline for each observation. Then we fix the RCP and LCP phase gains for the corresponding antenna to be zero \cite{Leppanen_1995_LPCAL}. We note that the phase gain prior with a range is analogous in part to setting up a reference antenna.

For millimeter-VLBI (mm-VLBI) data with short phase coherence time, we can use an uncorrelated normal-distribution $\phi \curvearrowleft \mathcal{N}(0, \sigma_{\phi}^{2})$ for the phase gain prior. In \texttt{resolve}, the temporal correlation structures in gain solutions are inferred from the data.
In other words, we determine the gain solution interval from the data without manual steering and we are even able to infer the solution interval per antenna, which can be advantageous for data from a heterogeneous VLBI array \citep{JSKIM_2025}. More details on antenna-based gain calibration in \texttt{resolve} can be found in \citep{Arras_2019, JSKim_2024}.

\paragraph{Generative model for D-term prior $D$}

The D-term (leakage) prior model is given by
\begin{align}
    D = \text{exp}(a + \mathbf{i} b),
\end{align}
where $a \curvearrowleft \mathcal{N}(m_{a}, \sigma_{a}^{2})$ and $b \curvearrowleft \mathcal{N}(0, \sigma_{b}^{2})$ are normal distributed priors, $m_{a}$ is the mean of the lognormal amplitude D-term, and $\sigma_{b}$ and $\sigma_{b}$ are the standard deviations of the lognormal amplitude and phase D-term, respectively.\\

Therefore, the D-term amplitude is lognormal distributed and the D-term phase is normal distributed. A summary table in \autoref{table:imaging_cal_prior} describes our calibration and imaging prior models.

\subsection{Posterior distribution} \label{sec:posterior}

The posterior distribution of all unknowns given the visibility data in \autoref{eq:Bayes' thm} is a very high dimensional object including the Stokes images, gains, and D-terms. Due to the high number of dimensions and the complicated relations between all the involved quantities, any representation of this probability function needs an approximation.
One way of representing a posterior distribution is via a set of samples drawn from it. If $\mathbf{s}=(\mathbf{I},\mathbf{G},\mathbf{D})$ denotes the quantities to be inferred and $\mathbf{d}=\textbf{V}$ denotes the data, then the samples $\mathbf{s}_i\hookleftarrow \mathcal{P}(\mathbf{s}|\mathbf{d})$ with $i\in\{1,\ldots N\}$ are an approximative representation of the posterior $\mathcal{P}(\mathbf{s}|\mathbf{d})$, as any expectation with respect to some function $f(\mathbf{s})$ can be calculated from those approximately
\begin{equation}
	\langle f(\mathbf{s}) \rangle_{(\mathbf{s}|\mathbf{d})}:= \int \!\! \mathcal{D}\mathbf{s}\, f(\mathbf{s}) \, \mathcal{P}(\mathbf{s}|\mathbf{d}) \approx \frac{1}{N} \sum_{i=1}^{N} f(\mathbf{s}_i), 
\end{equation}  

\noindent
where $\int\mathcal{D}\mathbf{s}$ indicates a path integral. \\
   
In this work, we utilized two variational inference algorithms (MGVI, geoVI) \citep{Knollmueller_2019_MGVI, Frank_2021_geoVI} as implemented in the \texttt{NIFTy} software package \citep{2013AandA...554A..26S, 2019ascl.soft03008A} to explore the posterior distribution of the Stokes images and calibration solutions. In the variational inference method, the posterior distribution is approximated as a parameterized distribution by minimizing the Kullback-Leibler (KL) divergence as a cost function. The KL divergence measures the information gain between two probability distributions. 

These variational inference methods scale quasi-linearly in computational complexity with the problem size; therefore, they enable the solution of high-dimensional calibration and imaging problems. Specifically, geoVI, an extension of the MGVI algorithm, can describe non-Gaussian posteriors by constructing a coordinate transformation between the latent space, in which the prior was Gaussian, but the posterior is not, and another latent space, in which the posterior becomes approximate Gaussian. More details about MGVI and geoVI can be found in \citep{Knollmueller_2019_MGVI, Frank_2021_geoVI}.

From the posterior samples, the posterior mean of the fractional linear polarization is 

\begin{equation}
    \langle P_{\text{frac}}(\mathbf{s}) \rangle_{(\mathbf{s}|\mathbf{d})} := \int \!\! \mathcal{D}\mathbf{s} \left( \frac{Q(\mathbf{s})^{2}+U(\mathbf{s})^{2}}{I(\mathbf{s})^{2}} \right) \mathcal{P}(\mathbf{s}|\mathbf{d}) = \left<\frac{Q^{2} + U^{2}}{I^{2}} \right>.
\end{equation}

We note that the posterior mean of the fractional linear polarization is not equal to $(\langle Q \rangle^{2} + \langle U \rangle^{2}) / \langle I \rangle ^{2}$, where $\langle I \rangle$, $\langle Q \rangle$, and $\langle U \rangle$ are the posterior mean Stokes images, because of the non-linear dependence of $P_{\text{frac}}$ on the Stokes parameters.

As another example, the posterior mean of the electric vector position angle (EVPA) is

\begin{equation}
    \langle \Phi \rangle_{(\mathbf{s}|\mathbf{d})} := \int \!\! \mathcal{D}\mathbf{s} \, \frac{1}{2} \,\text{arctan} \,{\left( \frac{U(\mathbf{s})}{Q(\mathbf{s})} \right)} \, \mathcal{P} (\mathbf{s}|\mathbf{d}) = \left< \frac{1}{2} \text{arctan} \left( \frac{U}{Q} \right)\right>.
\end{equation}

\begin{figure}
    \centering
    \includegraphics[width=9cm]{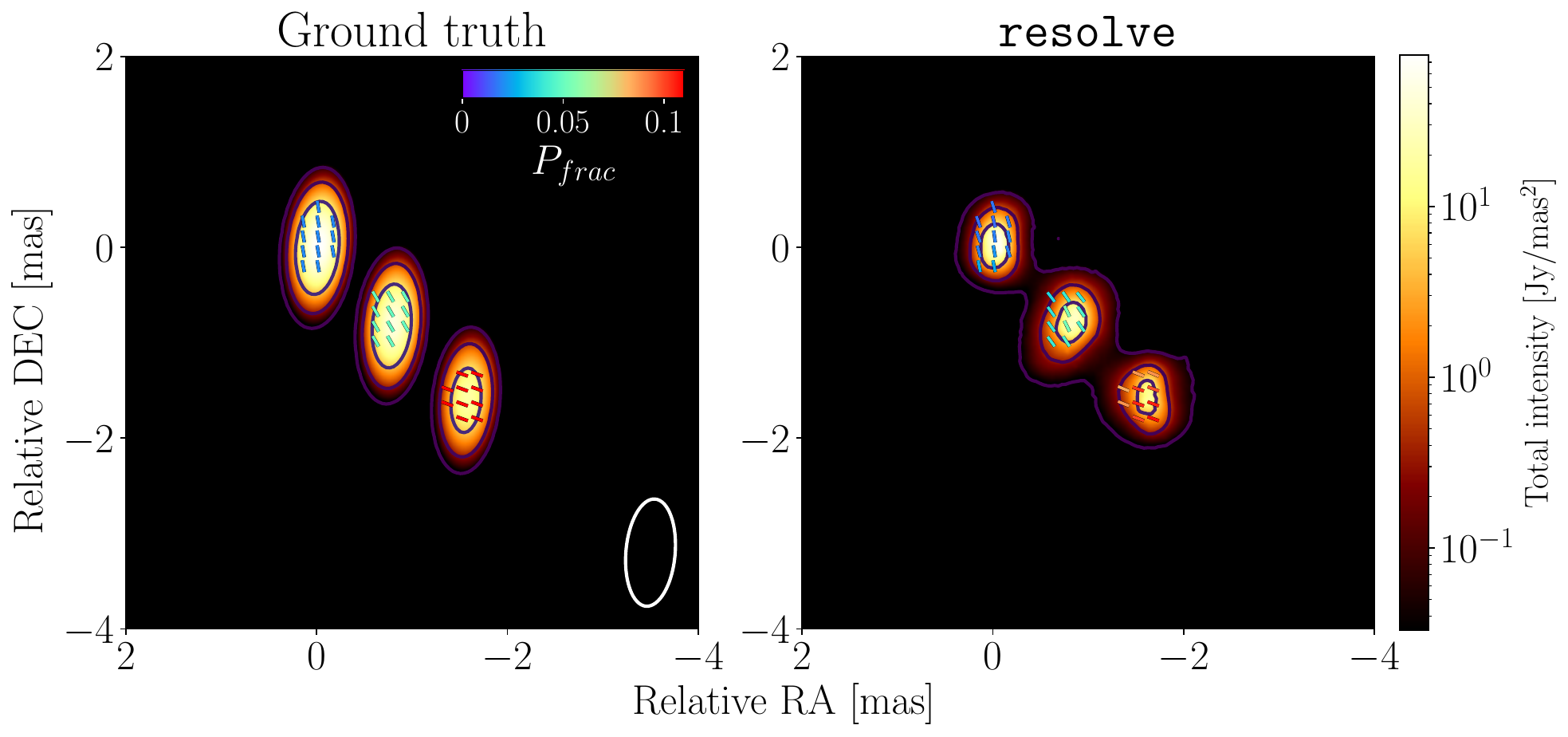}
    \caption{Comparison between the ground truth image (left panel) convolved with the nominal \texttt{CLEAN} beam with the uniform weighting and the posterior mean \texttt{resolve} polarization reconstruction (right panel) with EVPAs in all images with colors corresponding to the fractional linear polarization $P_{\text{frac}}$. The contours represent the total intensity of corresponding images.}\label{Synthetic_image}
\end{figure} 

\begin{figure}
    \centering
    \includegraphics[width=9cm]{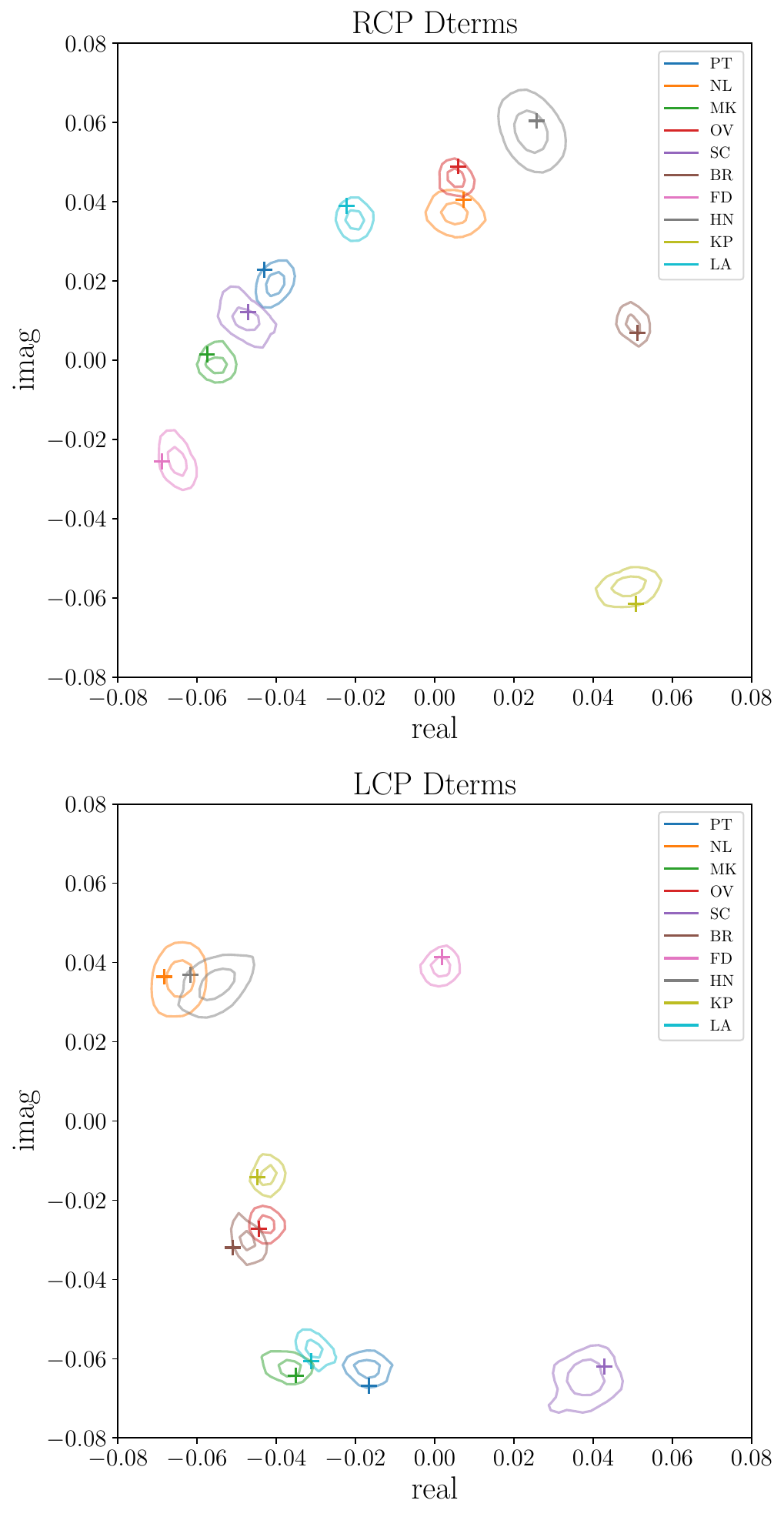}
    \caption{Comparison between the D-term posterior using \texttt{resolve} and ground truth D-terms from the synthetic data. Contours show 1$\sigma$ and 2$\sigma$ cumulative regions of \texttt{resolve} posterior D-terms using Gaussian kernel density estimation. The plus signs correspond to the ground truth D-terms.}\label{Synthetic_Dterm_plot}
\end{figure}

\section{Application to synthetic data}\label{Sect:synthetic_data}

In VLBI, polarization calibration using multiple calibrators at different declinations helps to reconstruct more robust leakage solutions, which break the degeneracy between the field rotation angle matrix $\mathbf{P}$ and the leakage matrix $\mathbf{D}$ \citep{Park_2021_GPCAL, Martividal_2021_PolSolve}. In order to validate the Bayesian polarization calibration method with multiple calibrators, we tested our method with three of the synthetic VLBA datasets presented in \citet{Park_2021_GPCAL}. Synthetic datasets were produced using \texttt{PolSimulate} in the \texttt{CASA} software \citep{CASA_2007} with OJ287, 3C273, BLLac UV-coverage at 15GHz by ten VLBA antennas. More details on synthetic data can be found in \citet{Park_2021_GPCAL}. The ground truth image consists of three point sources. The ground truth image convolved with the nominal \texttt{CLEAN} beam with uniform weighting is in \autoref{Synthetic_image}. We assume that there is no gain corruption in the data. 

Polarization calibration with synthetic data was performed as follows. First, initial D-term estimates were obtained using the maximum a posteriori (MAP) method using all three calibrator datasets. Then, the posterior distribution of Stokes images and D-term was reconstructed with 3C273 synthetic data using the geoVI method \citep{Frank_2021_geoVI} starting from the estimated MAP D-terms as the initial condition. \autoref{Synthetic_image} shows the \texttt{resolve} total intensity image and EVPAs with colors corresponding to the fractional linear polarization (right panel). We chose a spatial domain of 256 $\times$ 256 pixels and a field of view of 10 mas $\times$ 10mas. The \texttt{resolve} image was able to recover three linearly polarized components (with fractional linear polarization $2\%$, $5\%$, and $11\%$ respectively).

\autoref{Synthetic_Dterm_plot} represents a comparison between the \texttt{resolve} D-term posterior distribution and the ground truth D-terms. The \texttt{resolve} posterior distribution is obtained from 100 posterior samples using Gaussian kernel density distribution in \texttt{Scipy} Python library \citep{Scipy_2020}. Overall, the reconstructed D-term posterior distributions using \texttt{resolve} are consistent with the ground truth D-terms within the 2$\sigma$ errors. This result demonstrates that the \texttt{resolve} polarization calibration and imaging method is able to reconstruct reliable D-term solutions and Stokes images using multiple calibrator datasets. 

\begin{figure*}
    \centering
    \includegraphics[width=18cm]{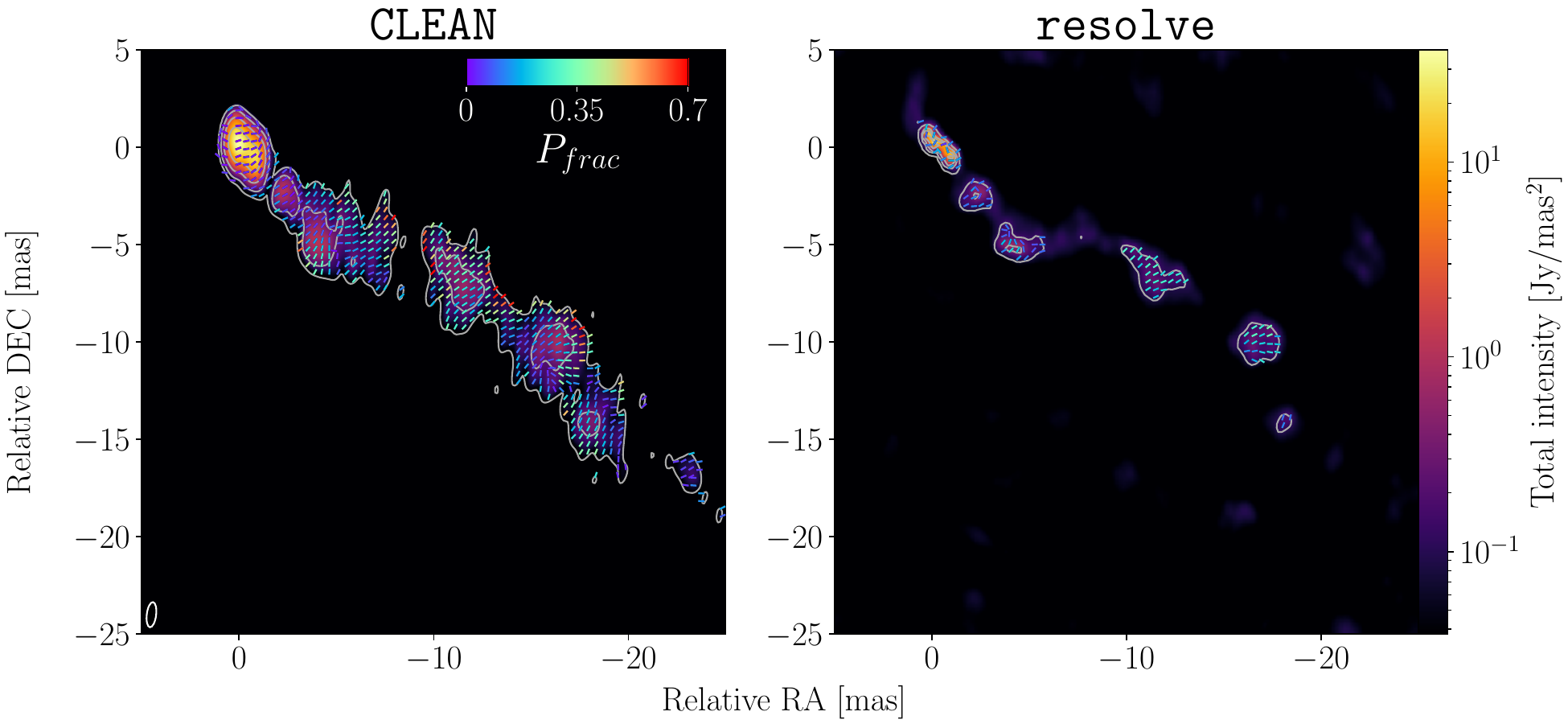}
    \caption{Comparison between the 3C273 VLBA \texttt{CLEAN} and \texttt{resolve} posterior mean linear polarization reconstructions at 15 GHz. Colored ticks indicate EVPAs in all images, with colors corresponding to the fractional linear polarization $P_{\text{frac}}$. The contours representing the total intensity of corresponding images increase by a factor of 4, starting from 0.3$\%$ of the peak total intensity of corresponding image.}\label{3C273_Linearpol}
\end{figure*}

\section{Application to real data}\label{Sect:real_data}

\subsection{3C273 VLBA observation at 15 GHz}

We applied our Bayesian polarization calibration and imaging method to precalibrated (without self-calibration and D-term calibration) 3C273 VLBA MOJAVE survey data at 15 GHz on January 28, 2017 \citep{Lister_2018_MOJAVE} to demonstrate that \texttt{resolve} is able to infer D-terms per intermediate frequency (IF) from a source with complex structure. The data have eight IFs (32MHz each) and the total bandwidth is 256MHz.

We reconstructed the \texttt{resolve} Stokes images with a spatial domain of 256 $\times$ 256 and a field of view of 50 mas $\times$ 50 mas. For the \texttt{resolve} reconstruction, we added a 10 percent systematic error budget in the data and selected Los Alamos (LA) as a reference antenna for the polarization calibration. The reduced $\chi^{2}$ of the \texttt{resolve} reconstruction was 1.45, and the number of posterior samples was 20. The wall-clock time for the \texttt{resolve} reconstruction was around 7 hours on a laptop with five message passing interface (MPI) tasks. 

We assumed that Stokes images do not vary over frequencies due to the narrow bandwidth (256 MHz), and performed antenna gain self-calibration and leakage calibration per IF. We assumed that each antenna has four different correlation kernels for the log-amplitude gain and phase gain and two polarization modes. The same correlation kernel is inferred over multi-IFs. The model parameters for the polarization imaging prior, gain prior, and D-term prior for the 3C273 VLBA data can be found in \autoref{table:model_parameters_polsky_3C273} and \autoref{table:model_parameters_cal_3C273}.

\autoref{3C273_Linearpol} shows a comparison of linear polarization reconstructions from \texttt{CLEAN} and the \texttt{resolve} posterior mean. The EVPA colors correspond to the fractional linear polarization, and contours represent the total intensity of the corresponding images. The \texttt{CLEAN} images are taken from the MOJAVE archive\footnote{\url{https://www.cv.nrao.edu/MOJAVE/}}. For the \texttt{CLEAN} reconstruction, the MOJAVE team performed self-calibration and image reconstruction using the \texttt{DIFMAP} software \citep{Shepherd_1997_DIFMAP} and the \texttt{LPCAL} task \citep{Leppanen_1995_LPCAL} in the \texttt{AIPS} software \citep{Greisen_2003_AIPS} for polarization calibration. The D-term solutions using \texttt{LPCAL} are the median solution values of all sources in the epoch after removing obvious outliers \citep{Lister_2005}.

In the core region, the \texttt{resolve} image (right panel) has a higher resolution compared with the core in the \texttt{CLEAN} image (left panel). We note that it is also possible to use an over-resolved \texttt{CLEAN} beam to obtain a higher-resolution restored image. The extended total intensity emission in the \texttt{resolve} image looks thinner than the \texttt{CLEAN} image. The jet emissions in the \texttt{resolve} image are overall consistent with the \texttt{CLEAN} reconstruction with bright jet emissions (see the contour from 1.2$\%$ of the peak total intensity).

The linear polarization comparison between \texttt{CLEAN} and \texttt{resolve} shows noticeable differences. The fractional linear polarization on the edges of linearly polarized emission in the \texttt{CLEAN} reconstruction is relatively higher than that in the \texttt{resolve} reconstruction. The discrepancy may result from the lack of a polarization constraint in the \texttt{CLEAN} imaging prior or biases from the similarity approximation in \texttt{CLEAN}-based polarization calibration. In contrast to \texttt{CLEAN}, \texttt{resolve} can describe complex source structure spanning a range of spatial scales using the Gaussian process sky prior we described, which is aware of spatial correlations of the polarized flux while ensuring the polarization constraints at each image pixel individually. Encoding the polarization constraint facilitates physically sensible reconstruction consistent with the theoretical fractional linear polarization limit (up to 75 $\%$) for optically thin synchrotron radiation from non-thermal electrons with a power-law distribution of energies \citep{Rybicki_1979}. 

The EVPA pattern is generally consistent except for the core region, due to the improved resolution in the \texttt{resolve} image. Both images show the rotational EVPA structure at the core, but the \texttt{resolve} image exhibits thinner emission structures that show more abrupt changes. \autoref{Fig:3C273_EVPA_std} shows the 3C273 \texttt{resolve} EVPA standard deviation. We note that a robust rotation measure can be estimated using the Bayesian approach \citep{Vogt_Ensslin_2005_Bayesian_RM}. A detailed investigation of Bayesian rotation measure analysis using our method is left for future work. In the \texttt{resolve} reconstruction, Stokes V emission is negligible (less than 0.2 $\%$ of the total intensity emission). The Stokes V reconstruction requires additional calibration steps. This aspect will be addressed in future work.

For the leakage calibration, a total of 160 D-terms (two polarization modes $\times$ ten antennas $\times$ eight IFs) are inferred since the D-terms in MOJAVE observations with VLBA at 15 GHz tend to be different per IF. This approach of reconstructing D-terms per IF is highly desirable, especially for wideband observations, as the signal path for each IF is different, which can result in changes to the D-terms for each IF separately. We also note that baseband boundaries can cause jumps in leakage that will also differ for individual IFs \citep{Martividal_2021_PolSolve}. 

In \autoref{3C273_15GHz_Dterm_cross_correlation}, the \texttt{resolve} D-term posterior means and the D-terms obtained using the \texttt{LPCAL} software are shown to be consistent with each other, as they are strongly correlated. It is important to note that the \texttt{resolve} D-term solutions are from 3C273 data only, and the D-terms using \texttt{LPCAL} are the median values of D-term solutions from multiple sources. The results indicate that our Bayesian polarization calibration method is able to estimate reliable D-terms from data with complex source structures. A more detailed analysis with additional datasets is deferred to future work.

\begin{figure*}[t]
    \centering
    \includegraphics[width=18cm]{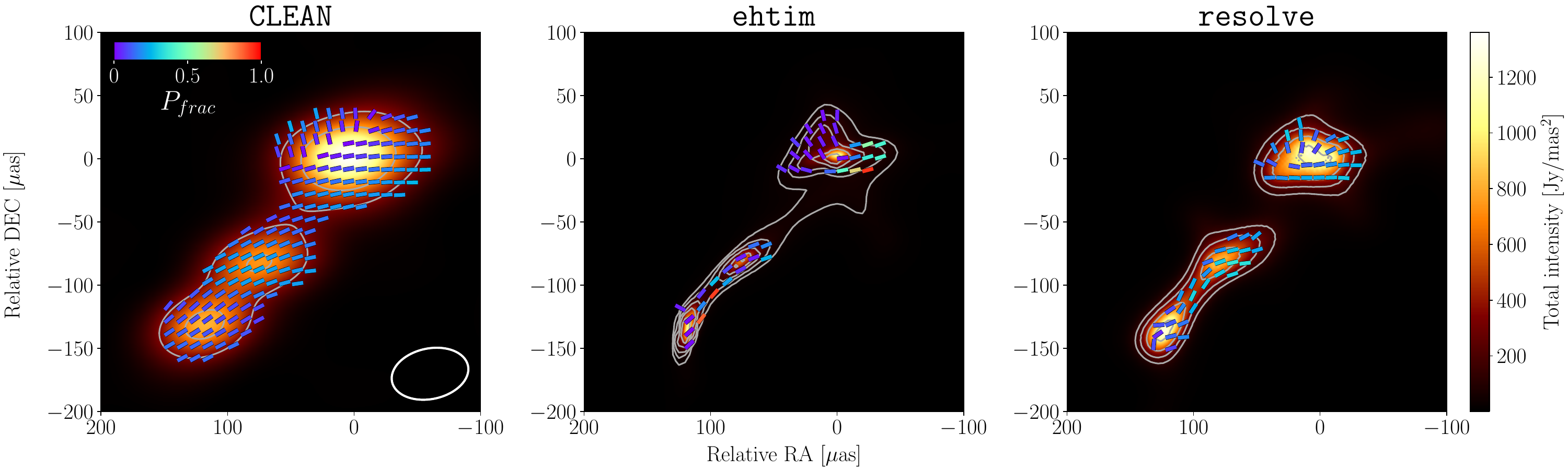}
    \caption{Comparison among the OJ287 GMVA+ALMA \texttt{CLEAN}, \texttt{ehtim}, and \texttt{resolve} posterior mean linear polarization reconstructions at 86 GHz. Colored ticks indicate EVPAs in all images, with colors corresponding to the fractional linear polarization $P_{\text{frac}}$. The contours representing the total intensity of the corresponding images increase by a factor of 2, starting from 10$\%$ of the peak \texttt{resolve} posterior mean total intensity. All images were processed by Gaussian interpolation.}\label{OJ287_Linearpol}
\end{figure*}

\subsection{OJ287 GMVA+ALMA observation at 86 GHz}\label{sec:OJ287_GMVA}

\citet{Zhao_2022} reported the GMVA+ALMA observation of the blazar OJ 287 at 86 GHz on April 2, 2017. The blazar OJ 287 is a supermassive binary black hole candidate that shows quasiperiodic optical outbursts with a period of about 12 years \citep{Sillanpaa_1988}. 
GMVA observations jointly with the phased ALMA provide high fidelity images due to improved sensitivity and long north-south baselines \citep{Issaoun_2019, Zhao_2022, Lu_2023, JSKIM_2025}. However, polarization calibration and imaging for mm-VLBI observations are demanding due to the low S/N of the polarized signals, tropospheric phase corruption, and heterogeneous antenna statistics, such as in GMVA+ALMA data.

We revisit the GMVA+ALMA observation of the blazar OJ 287 at 86 GHz in \citet{Zhao_2022} to validate our Bayesian polarization calibration and imaging method using precalibrated data. For polarization calibration, robust gain self-calibration is required due to the degeneracy between D-terms and gains. \cite{JSKim_2024, JSKIM_2025} validate the \texttt{resolve} self-calibration methods with mm-VLBI datasets. Reconstructed antenna gain solutions from the VLBA M87 observation in 2013 at 43GHz indicate that gain solutions from a homogeneous array with high S/N are consistent among \texttt{CLEAN}, \texttt{ehtim}, and \texttt{resolve} methods \citep{JSKim_2024}. However, \texttt{CLEAN} self-calibration utilizes a crude regularizer (e.g., a uniform solution interval for different antenna gain solutions) and often flags a significant fraction of the data. Therefore, these limitations may hinder robust self-calibration for mm-VLBI observations. 

To mitigate this issue, we employed different log-amplitude gain correlation kernels per antenna in order to take into account heterogeneous array sensitivity. For the gain phase prior, we utilized uncorrelated normally distributed phases due to the short phase coherence time in GMVA observations at 86 GHz \citep{JSKIM_2025}. More specifically, we did not model the temporal correlation in phase solutions, which is analogous to setting up phase solution intervals smaller than the averaged time for all antennas. Employing a different gain amplitude kernel per antenna is analogous to setting up different solution interval constraints per antenna in self-calibration. In \texttt{resolve}, the time correlations in the gain solutions are inferred from the data in an automated fashion, and the large uncertainties of particular data points are taken into account naturally in the Bayesian framework. Therefore, we are able to perform more robust self-calibration without flagging a significant number of data points compared to \texttt{CLEAN}-based self-calibration methods. Furthermore, the large gain uncertainties inherent to mm-VLBI observations are explicitly accounted for and propagated into the image domain. As a result, the reliability of the calibration and imaging is quantified through uncertainty estimation.

Polarization calibration for mm-VLBI datasets presents several additional challenges. In mm-VLBI, we often use the target data for D-term calibration due to a lack of point-like calibrators. Polarization leakages tend to be higher in mm-VLBI than in centimeter-VLBI, and the high resolution often reveals complex source polarization structures. Thus, precise D-term calibration is crucial to obtain reliable polarimetric images, and accurate polarimetric images, in turn, improve D-term calibration. However, in conventional polarization calibration methods, utilizing the similarity approximation or \texttt{CLEAN} images as a prior hinders the reconstruction of complex polarization structures since a few subcomponents in the similarity approximation and the collection of delta components in the \texttt{CLEAN} reconstruction cannot accurately represent the extended emission and complex source structures that are smaller than the \texttt{CLEAN} beam. On the other hand, in \texttt{resolve}, polarization imaging models with complex source structures, satisfying the polarization constraint and consistent with the data, are utilized for more robust polarization calibration. 

The data are time-averaged with 15 seconds and frequency-averaged for \texttt{ehtim} and \texttt{resolve} software, since the antenna leakages in the data are similar over frequencies \citep{Zhao_2022}. For the \texttt{CLEAN} image, self-calibration and image reconstruction were performed using the \texttt{DIFMAP} software and the \texttt{LPCAL} method of the \texttt{AIPS} software was used for leakage calibration using four individual intermediate frequencies (IFs). For the ehtim reconstruction, self-calibration and polarization calibration were performed in an iterative fashion using the \texttt{ehtim} software. The details regarding the CLEAN and ehtim polarization calibration and image reconstructions can be found in \cite{Zhao_2022}. 

For the \texttt{resolve} reconstruction, Stokes I, Q, U, and V images with a spatial domain of 256 $\times$ 256 pixels and a field of view of 500 $\mu$as $\times$ 500 $\mu$as, as well as gain solutions with a time interval of 15 seconds and leakage solutions, are inferred simultaneously. The reduced $\chi^{2}$ of the \texttt{resolve} reconstruction was 1.2, and the number of posterior samples was 100. The wall-clock time for the \texttt{resolve} reconstruction was around 8.5 hours on a single node of the MPIfR cluster with 25 MPI tasks. The model parameters for the polarization imaging prior, gain prior, and D-term prior for the OJ287 GMVA+ALMA data can be found in \autoref{table:model_parameters_polsky_OJ287} and \autoref{table:model_parameters_cal_OJ287}.

\autoref{OJ287_Linearpol} depicts the total intensity with EVPAs of the blazar OJ287 at 86 GHz using three different imaging algorithms: \texttt{CLEAN}, \texttt{ehtim} from \cite{Zhao_2022}, and \texttt{resolve}. All three image reconstructions show three main components in total intensity and diverging EVPAs in the northwest component. Furthermore, the curved jet (middle component) from the core (southeast component) is recognizable only in the \texttt{ehtim} and \texttt{resolve} images. In contrast, \texttt{CLEAN} has limitations in reconstructing smaller scale structures than the \texttt{CLEAN} beam. The \texttt{ehtim} and \texttt{resolve} images show better resolution than the \texttt{CLEAN} image because forward modeling permits partial removal of the observational point spread function during the data inversion process. The \texttt{ehtim} image exhibits higher resolution than the \texttt{resolve} image, since the sparsity promoting regularizers in \texttt{ehtim} tend to produce extremely sharp structures.

For polarization imaging, polarization constraints are encoded in \texttt{resolve} and \texttt{ehtim}. We performed the absolute EVPA calibration using the reported integrated EVPA of ALMA-only OJ287 image at 86 GHz in \cite{Goddi_2021}. The details of the absolute EVPA calibration method can be found in \autoref{sec:Absolute_EVPA_cal}. In \autoref{OJ287_Linearpol}, the EVPAs in the northwest jet component diverge in all three images. 
The fractional linear polarization along the northeast direction in the diverging jet component increases gradually in the images from \texttt{CLEAN} and \texttt{resolve}, whereas the \texttt{ehtim} image shows a nearly monotonic fractional linear polarization. In the \texttt{ehtim} image, there are extremely highly polarized pixels ($P_{\text{frac}}$ > 99 $\%$) that may be imaging artifacts caused by the sparsity regularizers.

The southeast core component and the middle jet component show EVPAs along the jet direction. This indicates toroidal magnetic fields, given the relatively small Faraday rotation ($3.05 \pm 0.62 \times 10^{3} \text{rad} \, \text{m}^{-2}$) from the ALMA observation \citep{Goddi_2021}. In the \texttt{ehtim} and \texttt{resolve} images, we see bending EVPA patterns, and the pattern is more wiggling in the \texttt{resolve} image. The diverging EVPA pattern in the northwest jet component can be explained by oblique or recollimation shocks \citep{Zhao_2022}. \autoref{Fig:OJ287_EVPA_std} shows the OJ287 \texttt{resolve} EVPA posterior standard deviation. In the \texttt{resolve} reconstruction, the OJ287 Stokes V emission is negligible (less than 0.1 $\%$ of the total intensity emission).

\begin{figure}
    \centering
    \includegraphics[width=9cm]{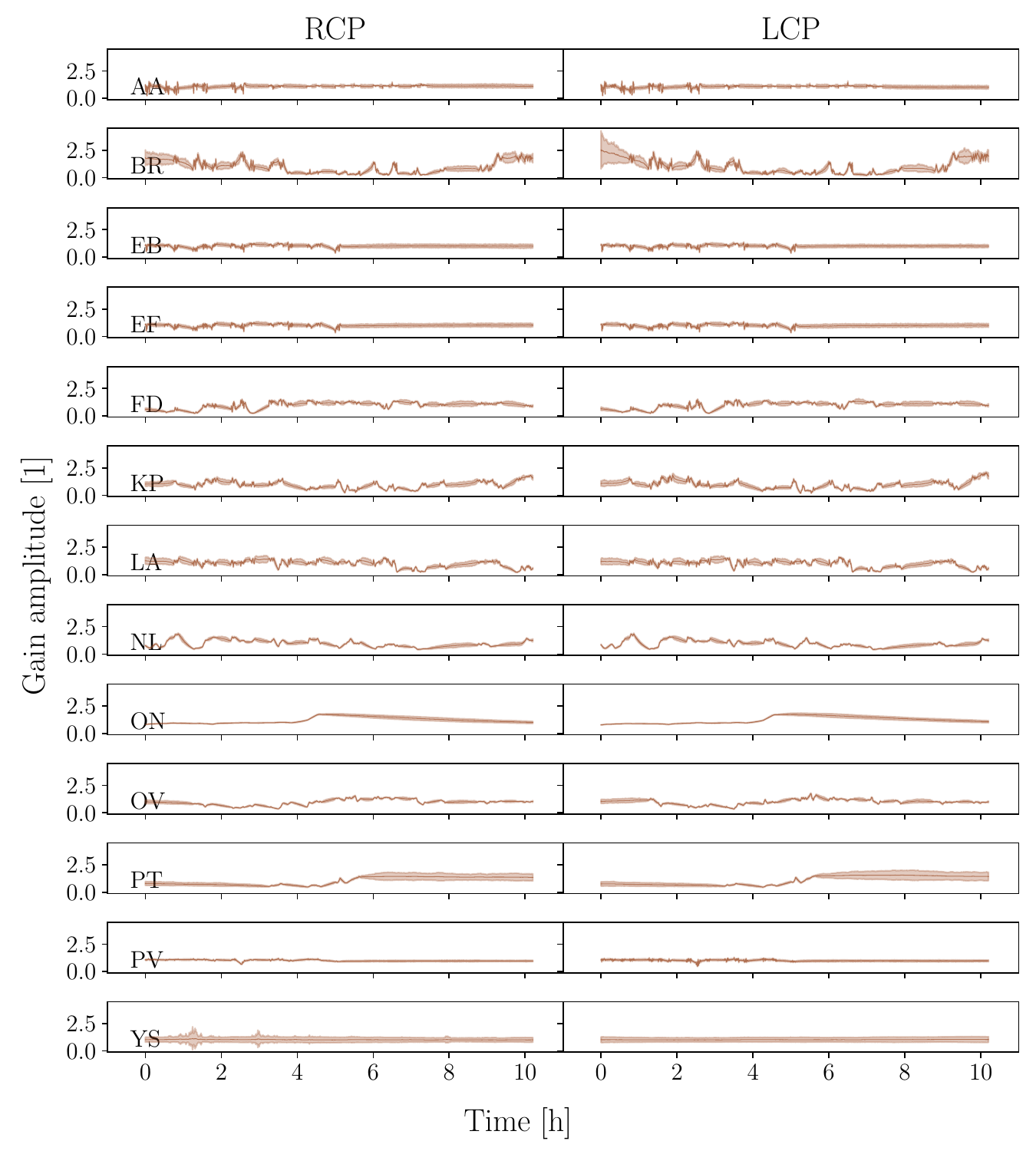}
    \caption{Posterior amplitude gains of the OJ287 GMVA+ALMA observation using \texttt{resolve}. The solid line represents the amplitude gain posterior mean with a semi-transparent standard deviation. Each row represents an individual antenna with corresponding abbreviated name in the bottom left corner of each RCP plot.} \label{OJ287_amp_gain}
\end{figure}

\begin{figure}
    \centering
    \includegraphics[width=9cm]{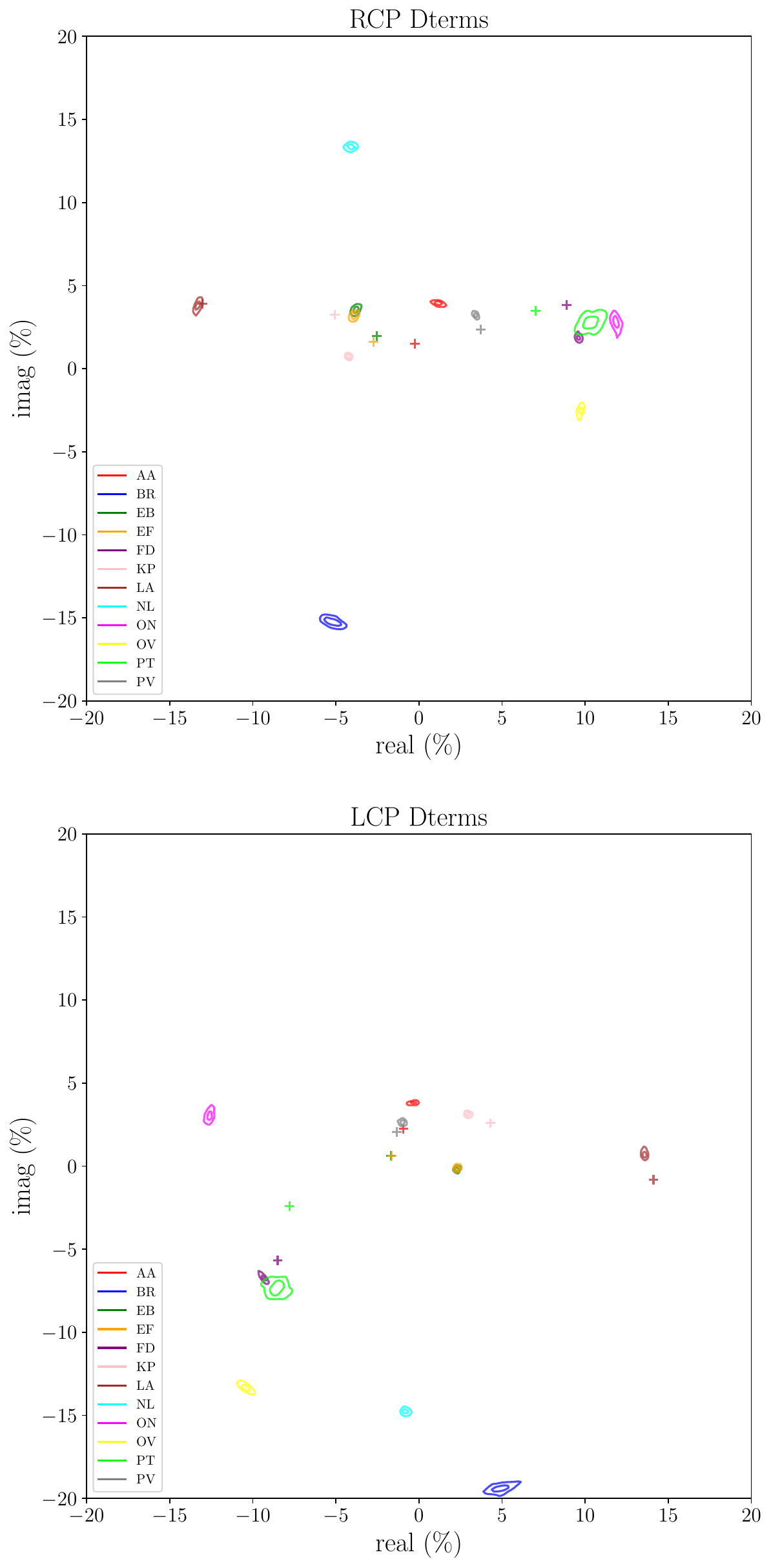}
    \caption{Comparison between OJ287 GMVA+ALMA D-term posterior distributions using \texttt{resolve} and D-term solutions using \texttt{ehtim}. Contours show 1$\sigma$ and 2$\sigma$ cumulative regions of \texttt{resolve} posterior D-terms using Gaussian kernel density estimation. The plus signs correspond to the reconstructed D-terms using \texttt{ehtim}.} \label{OJ287_Dterms}
\end{figure}

\autoref{OJ287_amp_gain} shows the amplitude gain posterior distribution. We employed separate correlation kernels per antenna to infer different temporal correlation structures arising from the heterogeneous antenna sensitivities. We utilized an uncorrelated phase gain prior since the phase coherence time is comparable to the averaging time (15 seconds). 

\autoref{OJ287_Dterms} depicts a comparison between the \texttt{resolve} D-term posterior distributions and the reconstructed D-terms using \texttt{ehtim}. The \texttt{resolve} and \texttt{ehtim} D-terms are broadly consistent. In \autoref{OJ287_Dterm_cross_correlation}, cross-correlation plots between the \texttt{resolve} posterior D-terms and \texttt{ehtim} reconstructed D-terms are shown. They exhibit strong positive correlations, but some antennas (e.g., PT) show weaker positive correlations. This discrepancy between \texttt{resolve} and \texttt{ehtim} results from the highly corrupted phases at 3mm and the different self-calibration and flagging routines used. For polarization calibration and imaging, \texttt{ehtim} used eight antennas out of 13, whereas \texttt{resolve} utilized all antennas (YS observed only a single polarization; therefore, we show the \texttt{resolve} D-terms for 12 antennas). In \texttt{resolve}, non-Gaussian leakage posterior distributions were obtained using the geoVI method. A few antennas (e.g., BR and ON) exhibit higher uncertainty in the D-term phase due to highly corrupted visibility phases. The \texttt{ehtim} D-terms are overall consistent with D-terms obtained from the \texttt{LPCAL} software \citep{Zhao_2022}. In conclusion, the D-term solutions are in good agreement across the \texttt{resolve}, \texttt{ehtim}, and \texttt{LPCAL} methods.

We note that regularized maximum likelihood (RML) methods can be interpreted as MAP estimation in Bayesian language \citep{JSKim_2024}. However, the point estimation cannot represent calibration solutions with high uncertainties well, and the MAP estimation is prone to overfitting the data. Furthermore, \texttt{ehtim} does not support multiscale base functions. In contrast to \texttt{resolve}, it tends to produce extremely sharp features. Different correlation structures between the core and the jet cannot be adequately described due to sparsity promoting priors, such as the $l_{1}$-norm, total variation (TV), and total squared variation (TSV). Employing extremely sharp images as a model in polarization calibration may result in biased D-terms and polarization images. A more detailed discussion of the Bayesian interpretation of regularizers in the RML method can be found in Appendix B of \citet{JSKim_2024}.

\subsection{Absolute EVPA calibration}\label{sec:Absolute_EVPA_cal}
Most VLBI observations use circular polarization feeds. Therefore, we need to constrain the arbitrary phase difference between the RCP and LCP feeds at each frequency in the fully station-based approach to the polarization classification and calibration \citep{Cotton_1993,Leppanen_1995_LPCAL}. This process is called absolute EVPA calibration. The simplest approach to determining this phase offset is to use a trusted external observation as an "anchor". External anchors are typically observations with a single dish, an interferometer with linear polarization feeds, or an observation that has already calibrated the phase difference between RCP and LCP independently. This method relies on the assumption that the signal of a "target" observation is similar to that of the anchoring observation, in both time and the physical structure of the emission. As such, the best calibrated sources for absolute EVPA calibration have the following three characteristics:
\begin{enumerate}
    \item The source has sufficient linear polarization in both the target and anchor observation.
    \item The source is not highly variable in time; as such it does not vary between the two observations.
    \item The source is compact; this is particularly relevant when comparing a target VLBI observation against single-dish or unresolved non-VLBI observations as we assume the emission from both observations comes from the same physical structure of the source. 
\end{enumerate}

To compute the RCP-LCP phase correction, we take twice the EVPA offset between the target and anchor observations. Typically, the EVPA offset is found by calculating the integrated Stokes Q and U flux density of both observations and calculating an integrated EVPA as $\mathrm{EVPA} = 0.5 \arctan[\mathrm{U}/\mathrm{Q}]$. This can be done by only considering Stokes Q and U emission that is co-spatial with Stokes I emission to ensure a direct comparison between the target and anchor.

In the MOJAVE survey, the absolute EVPA calibration relied on optically thin jet features in multiple sources that have relatively stable EVPAs. Based on the comparisons of compact AGNs with near-simultaneous single-dish observations, the MOJAVE team estimated that the VLBA EVPA measurements are accurate to $\sim 5 \%$\citep{Lister_2018_MOJAVE, Lister_2021}. For our reconstruction of 3C273 in \autoref{3C273_Linearpol}, we applied the same correction value from the MOJAVE team for the \texttt{CLEAN} reconstruction. For our reconstruction of OJ287 in \autoref{OJ287_Linearpol}, we calculated the integrated EVPA of our reconstructed \texttt{resolve} image of Stokes Q and U and compared it with ALMA observations with the EVPA - 69.69$^\circ$ at 86.3 GHz in \citet{Goddi_2021}. From this we found an EVPA correction of -16.07$^\circ$ based on the OJ287 \texttt{resolve}-integrated EVPA and ALMA-only EVPA.

In short, the \texttt{resolve} reconstructions of 3C273 at 15 GHz OJ 287 at 86 GHz demonstrate that our Bayesian polarization calibration and imaging pipeline can obtain reliable super-resolution polarimetric images, D-terms, and gain solutions from VLBI data. Reconstructed Stokes images are utilized as a polarization calibration model, enabling us to obtain more complex polarization structures than conventional \texttt{CLEAN}-based methods. Uncertainty estimation of reconstructed images and calibration solutions in \texttt{resolve} is especially beneficial in mm-VLBI due to the high calibration uncertainties from tropospheric phase corruptions and antenna leakages. In our method, calibration uncertainties are accounted for in our reconstructed images, and we are able to quantify the reliability of Stokes images and calibration solutions. The \texttt{resolve} reconstruction FITS files and results are available in the Zenodo archive \footnote{\url{https://zenodo.org/records/17699222}}.


\section{Conclusions}\label{Sect:conclusion}

In this work, we have presented a novel Bayesian polarization calibration and imaging method using the \texttt{resolve} algorithm. Our method can simultaneously reconstruct high-resolution polarimetric images that satisfy the polarization constraint ($I \ge  \sqrt{Q^{2}+U^{2}+V^{2}}$), antenna-based gains, and polarization leakages from the entire dataset. The polarization calibration is based on the reconstructed Stokes images instead of the similarity approximation used in traditional approaches, which does not hold for complex polarized emission patterns. Therefore, we can reconstruct reliable polarimetric images with complex structures and more robust calibration solutions, consistent with the entire dataset. Moreover, multisource and multi-IF polarization calibration is supported due to the scalability of \texttt{resolve}, achieved through variational inference methods \citep{Knollmueller_2019_MGVI, Frank_2021_geoVI}. 

We have demonstrated our method with synthetic and real observations. Examples using the quasar 3C273 MOJAVE VLBA data at 15 GHz and the blazar OJ287 GMVA+ALMA observation at 86 GHz show that \texttt{resolve} can reconstruct polarimetric images with super-resolution \citep{Honma_2014}, revealing structures below the synthesized beam, which is the resolution limit of conventional \texttt{CLEAN}-based methods. The fractional linear polarization in \texttt{resolve} images is consistent with the theoretical estimation of synchrotron radiation (< 75$\%$) in AGN jets. 

Reconstructed D-terms obtained with \texttt{resolve} are consistent with those from the conventional \texttt{CLEAN}-based method and the RML-based method \texttt{ehtim}. Furthermore, our RIME-based Bayesian polarization calibration and imaging method is able to reconstruct D-term solutions from target data with complex source structure without using calibrators.

Validation with synthetic and real datasets indicates that incorporating polarization calibration into imaging is beneficial for sparse and noisy VLBI datasets in order to estimate reliable and reproducible calibration solutions and images. Calibration uncertainties are explicitly taken into account in our final \texttt{resolve} results, and the reliability of the gain and leakage solutions was estimated. For future work, the EVPA posterior distribution can be utilized for rotation measure analysis, and frequency-dependent D-term calibration within the Bayesian framework should be investigated for wideband radio interferometric observations. 

\section*{Data availability}
The pipeline is publicly accessible at \url{https://github.com/JongseoKim/resolve_polcal}

\begin{acknowledgements}
      We thank the anonymous referee for constructive comments and suggestions, Yuri Kovalev for comments on the manuscript, Guang-Yao Zhao and Jose Gomez for providing the GMVA+ALMA OJ287 data, MOJAVE team for providing the VLBA 3C273 data. J. K. received financial support for this research from the International Max Planck Research School (IMPRS) for Astronomy and Astrophysics at the Universities of Bonn and Cologne. This work was supported by the M2FINDERS project funded by the European Research Council (ERC) under the European Union's Horizon 2020 Research and Innovation Programme (Grant Agreement No. 101018682).
      J. R. acknowledges financial support from the German Federal Ministry of Education and Research (BMBF) under grant 05A23WO1 (Verbundprojekt D-MeerKAT III). This research has made use of data obtained with the Global Millimeter VLBI Array (GMVA), which consists of telescopes operated by the MPIfR, IRAM, Onsala, Metsahovi, Yebes, the Korean VLBI Network, the Greenland Telescope, the Green Bank Observatory, and the Very Long Baseline Array (VLBA). The VLBA and the GBT are facilities of the National Science Foundation operated under cooperative agreement by Associated Universities, Inc. The data were correlated at the correlator of the MPIfR in Bonn, Germany. This research has made use of data from the MOJAVE database that is maintained by the MOJAVE team \cite{Lister_2018_MOJAVE}.
\end{acknowledgements}

\bibliographystyle{aa}
\bibliography{reference}{}

@ARTICLE{Pushkarev_2023,
       author = {{Pushkarev}, A.~B. and {Aller}, H.~D. and {Aller}, M.~F. and {Homan}, D.~C. and {Kovalev}, Y.~Y. and {Lister}, M.~L. and {Pashchenko}, I.~N. and {Savolainen}, T. and {Zobnina}, D.~I.},
        title = "{MOJAVE - XX. Persistent linear polarization structure in parsec-scale AGN jets}",
      journal = {\mnras},
         year = 2023,
        month = apr,
       volume = {520},
       number = {4},
        pages = {6053-6069},
          doi = {10.1093/mnras/stad525},
archivePrefix = {arXiv},
       eprint = {2209.04842},
 primaryClass = {astro-ph.HE},
       adsurl = {https://ui.adsabs.harvard.edu/abs/2023MNRAS.520.6053P}
}

@ARTICLE{Lister_2005,
       author = {{Lister}, M.~L. and {Homan}, D.~C.},
        title = "{MOJAVE: Monitoring of Jets in Active Galactic Nuclei with VLBA Experiments. I. First-Epoch 15 GHz Linear Polarization Images}",
      journal = {\aj},
         year = 2005,
        month = oct,
       volume = {130},
       number = {4},
        pages = {1389-1417},
          doi = {10.1086/432969},
archivePrefix = {arXiv},
       eprint = {astro-ph/0503152},
 primaryClass = {astro-ph},
       adsurl = {https://ui.adsabs.harvard.edu/abs/2005AJ....130.1389L}
}

@ARTICLE{Gabuzda_2017,
       author = {{Gabuzda}, D.~C. and {Roche}, N. and {Kirwan}, A. and {Knuettel}, S. and {Nagle}, M. and {Houston}, C.},
        title = "{Parsec scale Faraday-rotation structure across the jets of nine active galactic nuclei}",
      journal = {\mnras},
         year = 2017,
        month = dec,
       volume = {472},
       number = {2},
        pages = {1792-1801},
          doi = {10.1093/mnras/stx2127},
archivePrefix = {arXiv},
       eprint = {1709.09062},
 primaryClass = {astro-ph.GA},
       adsurl = {https://ui.adsabs.harvard.edu/abs/2017MNRAS.472.1792G}
}

@ARTICLE{Hovatta_2012,
       author = {{Hovatta}, Talvikki and {Lister}, Matthew L. and {Aller}, Margo F. and {Aller}, Hugh D. and {Homan}, Daniel C. and {Kovalev}, Yuri Y. and {Pushkarev}, Alexander B. and {Savolainen}, Tuomas},
        title = "{MOJAVE: Monitoring of Jets in Active Galactic Nuclei with VLBA Experiments. VIII. Faraday Rotation in Parsec-scale AGN Jets}",
      journal = {\aj},
         year = 2012,
        month = oct,
       volume = {144},
       number = {4},
          eid = {105},
        pages = {105},
          doi = {10.1088/0004-6256/144/4/105},
archivePrefix = {arXiv},
       eprint = {1205.6746},
 primaryClass = {astro-ph.CO},
       adsurl = {https://ui.adsabs.harvard.edu/abs/2012AJ....144..105H}
}

@article{Cotton_1993,
    pages = {1241},
    doi = {10.1086/116723},
    adsurl = {https://ui.adsabs.harvard.edu/abs/1993AJ....106.1241C},
    author = {{Cotton}, W.~D.},
    title = {{Calibration and Imaging OG Polarization Sensitive Very Long Baseline Interferometer Observations}},
    month = {September},
    volume = {106},
    year = {1993},
    journal = {\aj},
}

@article{Park_2023_GPCAL,
    volume = {958},
    title = {{Calibrating VLBI Polarization Data Using GPCAL. I. Frequency-dependent Calibration}},
    doi = {10.3847/1538-4357/acfd2f},
    archiveprefix = {arXiv},
    journal = {\apj},
    month = {November},
    primaryclass = {astro-ph.IM},
    eprint = {2310.03242},
    number = {1},
    author = {{Park}, Jongho and {Asada}, Keiichi and {Byun}, Do-Young},
    year = {2023},
    eid = {27},
    pages = {27},
    adsurl = {https://ui.adsabs.harvard.edu/abs/2023ApJ...958...27P},
}

@ARTICLE{Park_2023_GPCAL2,
       author = {{Park}, Jongho and {Asada}, Keiichi and {Byun}, Do-Young},
        title = "{Calibrating VLBI Polarization Data Using GPCAL. II. Time-dependent Calibration}",
      journal = {\apj},
         year = 2023,
        month = nov,
       volume = {958},
       number = {1},
          eid = {28},
        pages = {28},
          doi = {10.3847/1538-4357/acfd30},
archivePrefix = {arXiv},
       eprint = {2310.03244},
 primaryClass = {astro-ph.IM},
       adsurl = {https://ui.adsabs.harvard.edu/abs/2023ApJ...958...28P}
}

@article{Leppanen_1995_LPCAL,
    year = {1995},
    month = {November},
    author = {{Leppanen}, K.~J. and {Zensus}, J.~A. and {Diamond}, P.~J.},
    volume = {110},
    title = {{Linear Polarization Imaging with Very Long Baseline Interferometry at High Frequencies}},
    journal = {\aj},
    pages = {2479},
    doi = {10.1086/117706},
    adsurl = {https://ui.adsabs.harvard.edu/abs/1995AJ....110.2479L},
}

@article{Martividal_2021_PolSolve,
    author = {{Mart{\'\i}-Vidal}, I. and {Mus}, A. and {Janssen}, M. and {de Vicente}, P. and {Gonz{\'a}lez}, J.},
    title = {{Polarization calibration techniques for the new-generation VLBI}},
    year = {2021},
    eprint = {2012.05581},
    month = {February},
    doi = {10.1051/0004-6361/202039527},
    primaryclass = {astro-ph.IM},
    journal = {\aap},
    eid = {A52},
    archiveprefix = {arXiv},
    pages = {A52},
    volume = {646},
    adsurl = {https://ui.adsabs.harvard.edu/abs/2021A&A...646A..52M},
}

@article{Park_2021_GPCAL,
    pages = {85},
    title = {{GPCAL: A Generalized Calibration Pipeline for Instrumental Polarization in VLBI Data}},
    volume = {906},
    journal = {\apj},
    archiveprefix = {arXiv},
    adsurl = {https://ui.adsabs.harvard.edu/abs/2021ApJ...906...85P},
    eprint = {2011.09713},
    number = {2},
    author = {{Park}, Jongho and {Byun}, Do-Young and {Asada}, Keiichi and {Yun}, Youngjoo},
    year = {2021},
    doi = {10.3847/1538-4357/abcc6e},
    eid = {85},
    month = {January},
    primaryclass = {astro-ph.IM},
}

@article{Park_2021_M87,
    year = {2021},
    doi = {10.3847/1538-4357/ac26bf},
    pages = {180},
    adsurl = {https://ui.adsabs.harvard.edu/abs/2021ApJ...922..180P},
    number = {2},
    journal = {\apj},
    primaryclass = {astro-ph.HE},
    eprint = {2107.13243},
    title = {{A Revised View of the Linear Polarization in the Subparsec Core of M87 at 7 mm}},
    author = {{Park}, Jongho and {Asada}, Keiichi and {Nakamura}, Masanori and {Kino}, Motoki and {Pu}, Hung-Yi and {Hada}, Kazuhiro and {Kravchenko}, Evgeniya V. and {Giroletti}, Marcello},
    volume = {922},
    month = {December},
    eid = {180},
    archiveprefix = {arXiv},
}

@article{Hamaker_1996,
    year = {1996},
    journal = {\aaps},
    pages = {137-147},
    title = {{Understanding radio polarimetry. I. Mathematical foundations.}},
    volume = {117},
    adsurl = {https://ui.adsabs.harvard.edu/abs/1996A&AS..117..137H},
    month = {May},
    author = {{Hamaker}, J.~P. and {Bregman}, J.~D. and {Sault}, R.~J.},
}

@article{Smirnov_2011,
    volume = {527},
    month = {March},
    adsurl = {https://ui.adsabs.harvard.edu/abs/2011A&A...527A.106S},
    doi = {10.1051/0004-6361/201016082},
    archiveprefix = {arXiv},
    pages = {A106},
    title = {{Revisiting the radio interferometer measurement equation. I. A full-sky Jones formalism}},
    year = {2011},
    eid = {A106},
    journal = {\aap},
    primaryclass = {astro-ph.IM},
    author = {{Smirnov}, O.~M.},
    eprint = {1101.1764},
}

@article{Jones_1941,
    doi = {10.1364/JOSA.31.000488},
    number = {7},
    author = {{Jones}, R. Clark},
    adsurl = {https://ui.adsabs.harvard.edu/abs/1941JOSA...31..488J},
    month = {July},
    title = {{New calculus for the treatment of optical systems. I. Description and discussion of the calculus}},
    volume = {31},
    journal = {Journal of the Optical Society of America (1917-1983)},
    pages = {488},
    year = {1941},
}

@article{Zhao_2022,
    title = {{Unraveling the Innermost Jet Structure of OJ 287 with the First GMVA + ALMA Observations}},
    author = {{Zhao}, Guang-Yao and {G{\'o}mez}, Jos{\'e} L. and {Fuentes}, Antonio and {Krichbaum}, Thomas P. and {Traianou}, Efthalia and {Lico}, Rocco and {Cho}, Ilje and {Ros}, Eduardo and {Komossa}, S. and {Akiyama}, Kazunori and et al.},
    doi = {10.3847/1538-4357/ac6b9c},
    primaryclass = {astro-ph.HE},
    archiveprefix = {arXiv},
    year = {2022},
    pages = {72},
    month = {June},
    number = {1},
    adsurl = {https://ui.adsabs.harvard.edu/abs/2022ApJ...932...72Z},
    eid = {72},
    journal = {\apj},
    volume = {932},
    eprint = {2205.00554},
}

@article{Issaoun_2019,
    archiveprefix = {arXiv},
    adsurl = {https://ui.adsabs.harvard.edu/abs/2019ApJ...871...30I},
    eid = {30},
    eprint = {1901.06226},
    journal = {\apj},
    year = {2019},
    pages = {30},
    number = {1},
    title = {{The Size, Shape, and Scattering of Sagittarius A* at 86 GHz: First VLBI with ALMA}},
    doi = {10.3847/1538-4357/aaf732},
    primaryclass = {astro-ph.HE},
    month = {January},
    volume = {871},
    author = {{Issaoun}, S. and {Johnson}, M.~D. and {Blackburn}, L. and {Brinkerink}, C.~D. and {Mo{\'s}cibrodzka}, M. and {Chael}, A. and {Goddi}, C. and {Mart{\'\i}-Vidal}, I. and {Wagner}, J. and {Doeleman}, S.~S. and et al.},
}

@article{Lu_2023,
    eprint = {2304.13252},
    doi = {10.1038/s41586-023-05843-w},
    adsurl = {https://ui.adsabs.harvard.edu/abs/2023Natur.616..686L},
    month = {April},
    number = {7958},
    journal = {\nat},
    year = {2023},
    title = {{A ring-like accretion structure in M87 connecting its black hole and jet}},
    archiveprefix = {arXiv},
    volume = {616},
    primaryclass = {astro-ph.HE},
    author = {{Lu}, Ru-Sen and {Asada}, Keiichi and {Krichbaum}, Thomas P. and {Park}, Jongho and {Tazaki}, Fumie and {Pu}, Hung-Yi and {Nakamura}, Masanori and {Lobanov}, Andrei and {Hada}, Kazuhiro and {Akiyama}, Kazunori and et al.},
    pages = {686-690},
}

@article{Lister_2018_MOJAVE,
    journal = {\apjs},
    year = {2018},
    doi = {10.3847/1538-4365/aa9c44},
    adsurl = {https://ui.adsabs.harvard.edu/abs/2018ApJS..234...12L},
    eid = {12},
    volume = {234},
    eprint = {1711.07802},
    number = {1},
    title = {{MOJAVE. XV. VLBA 15 GHz Total Intensity and Polarization Maps of 437 Parsec-scale AGN Jets from 1996 to 2017}},
    primaryclass = {astro-ph.GA},
    pages = {12},
    archiveprefix = {arXiv},
    author = {{Lister}, M.~L. and {Aller}, M.~F. and {Aller}, H.~D. and {Hodge}, M.~A. and {Homan}, D.~C. and {Kovalev}, Y.~Y. and {Pushkarev}, A.~B. and {Savolainen}, T.},
    month = {January},
}

@inproceedings{Greisen_2003_AIPS,
    year = {2003},
    pages = {109},
    series = {Astrophysics and Space Science Library},
    volume = {285},
    doi = {10.1007/0-306-48080-8_7},
    editor = {{Heck}, Andr{\'e}},
    adsurl = {https://ui.adsabs.harvard.edu/abs/2003ASSL..285..109G},
    author = {{Greisen}, E.~W.},
    month = {March},
    title = {{AIPS, the VLA, and the VLBA}},
    booktitle = {Information Handling in Astronomy - Historical Vistas},
}

@book{Thompson_2017,
    doi = {10.1007/978-3-319-44431-4},
    adsurl = {https://ui.adsabs.harvard.edu/abs/2017isra.book.....T},
    author = {{Thompson}, A. Richard and {Moran}, James M. and {Swenson}, Jr., George W.},
    title = {{Interferometry and Synthesis in Radio Astronomy, 3rd Edition}},
    year = {2017},
}

@article{Goddi_2021,
    eid = {L14},
    volume = {910},
    adsurl = {https://ui.adsabs.harvard.edu/abs/2021ApJ...910L..14G},
    year = {2021},
    title = {{Polarimetric Properties of Event Horizon Telescope Targets from ALMA}},
    number = {1},
    archiveprefix = {arXiv},
    month = {March},
    eprint = {2105.02272},
    pages = {L14},
    author = {{Goddi}, Ciriaco and {Mart{\'\i}-Vidal}, Iv{\'a}n and {Messias}, Hugo and {Bower}, Geoffrey C. and {Broderick}, Avery E. and {Dexter}, Jason and {Marrone}, Daniel P. and {Moscibrodzka}, Monika and {Nagai}, Hiroshi and {Algaba}, Juan Carlos and et al.},
    doi = {10.3847/2041-8213/abee6a},
    primaryclass = {astro-ph.GA},
    journal = {\apjl},
}

@article{Scipy_2020,
    title = {{{SciPy} 1.0: Fundamental Algorithms for Scientific
            Computing in Python}},
    journal = {Nature Methods},
    pages = {261--272},
    author = {Virtanen, Pauli and Gommers, Ralf and Oliphant, Travis E. and
            Haberland, Matt and Reddy, Tyler and Cournapeau, David and
            Burovski, Evgeni and Peterson, Pearu and Weckesser, Warren and
            Bright, Jonathan and {van der Walt}, St{\'e}fan J. and
            Brett, Matthew and Wilson, Joshua and Millman, K. Jarrod and
            Mayorov, Nikolay and Nelson, Andrew R. J. and Jones, Eric and
            Kern, Robert and Larson, Eric and Carey, C J and
            Polat, {\.I}lhan and Feng, Yu and Moore, Eric W. and
            {VanderPlas}, Jake and Laxalde, Denis and Perktold, Josef and
            Cimrman, Robert and Henriksen, Ian and Quintero, E. A. and
            Harris, Charles R. and Archibald, Anne M. and
            Ribeiro, Ant{\^o}nio H. and Pedregosa, Fabian and
            {van Mulbregt}, Paul and {SciPy 1.0 Contributors}},
    adsurl = {https://rdcu.be/b08Wh},
    doi = {10.1038/s41592-019-0686-2},
    volume = {17},
    year = {2020},
}

@inproceedings{CASA_2007,
    adsurl = {https://ui.adsabs.harvard.edu/abs/2007ASPC..376..127M},
    volume = {376},
    author = {{McMullin}, J.~P. and {Waters}, B. and {Schiebel}, D. and {Young}, W. and {Golap}, K.},
    title = {{CASA Architecture and Applications}},
    editor = {{Shaw}, R.~A. and {Hill}, F. and {Bell}, D.~J.},
    series = {Astronomical Society of the Pacific Conference Series},
    month = {October},
    booktitle = {Astronomical Data Analysis Software and Systems XVI},
    pages = {127},
    year = {2007},
}

@article{Asada_2008_3C273,
    doi = {10.1086/524000},
    month = {March},
    journal = {\apj},
    title = {{Time Variation of the Rotation Measure Gradient in the 3C 273 Jet}},
    archiveprefix = {arXiv},
    year = {2008},
    number = {1},
    volume = {675},
    pages = {79-82},
    author = {{Asada}, Keiichi and {Inoue}, Makoto and {Kameno}, Seiji and {Nagai}, Hiroshi},
    eprint = {0806.4231},
    adsurl = {https://ui.adsabs.harvard.edu/abs/2008ApJ...675...79A},
    primaryclass = {astro-ph},
}

@article{Arras_2019,
    month = {July},
    title = {{Unified radio interferometric calibration and imaging with joint uncertainty quantification}},
    pages = {A134},
    archiveprefix = {arXiv},
    primaryclass = {astro-ph.IM},
    eprint = {1903.11169},
    journal = {\aap},
    author = {{Arras}, Philipp and {Frank}, Philipp and {Leike}, Reimar and {Westermann}, R{\"u}diger and {En{\ss}lin}, Torsten A.},
    volume = {627},
    eid = {A134},
    year = {2019},
    adsurl = {https://ui.adsabs.harvard.edu/abs/2019A&A...627A.134A},
    doi = {10.1051/0004-6361/201935555},
}

@article{JSKim_2024,
    year = {2024},
    primaryclass = {astro-ph.IM},
    title = {{Bayesian self-calibration and imaging in very long baseline interferometry}},
    eid = {A129},
    month = {October},
    author = {{Kim}, Jong-Seo and {Nikonov}, Aleksei S. and {Roth}, Jakob and {En{\ss}lin}, Torsten A. and {Janssen}, Michael and {Arras}, Philipp and {M{\"u}ller}, Hendrik and {Lobanov}, Andrei P.},
    journal = {\aap},
    volume = {690},
    pages = {A129},
    eprint = {2407.14873},
    archiveprefix = {arXiv},
    adsurl = {https://ui.adsabs.harvard.edu/abs/2024A&A...690A.129K},
    doi = {10.1051/0004-6361/202449663},
}

@article{2013AandA...554A..26S,
    volume = {554},
    eid = {A26},
    year = {2013},
    doi = {10.1051/0004-6361/201321236},
    pages = {A26},
    adsurl = {https://ui.adsabs.harvard.edu/abs/2013AandA...554A..26S},
    primaryclass = {astro-ph.IM},
    journal = {\aap},
    archiveprefix = {arXiv},
    author = {{Selig}, M. and {Bell}, M.~R. and {Junklewitz}, H. and {Oppermann}, N. and {Reinecke}, M. and {Greiner}, M. and {Pachajoa}, C. and {En{\ss}lin}, T.},
    title = {{NIFTY - Numerical Information Field Theory. A versatile PYTHON library for signal inference}},
    month = {June},
    eprint = {1301.4499},
}

@misc{2019ascl.soft03008A,
    title = {{NIFTy5: Numerical Information Field Theory v5}},
    eprint = {1903.008},
    adsurl = {https://ui.adsabs.harvard.edu/abs/2019ascl.soft03008A},
    pages = {ascl:1903.008},
    author = {{Arras}, P. and {Baltac}, M. and {En{\ss}lin}, T. and {Frank}, P. and {Hutschenreuter}, S. and {Knollmueller}, J. and {Leike}, R. and {Newrzella}, M. and {Platz}, L. and {Reinecke}, M. and {Stadler}, J.},
    eid = {ascl:1903.008},
    archiveprefix = {ascl},
    year = {2019},
    month = {March},
}

@ARTICLE{Roth_24,
       author = {{Roth}, Jakob and {Frank}, Philipp and {Bester}, Hertzog L. and {Smirnov}, Oleg M. and {Westermann}, R{\"u}diger and {En{\ss}lin}, Torsten A.},
        title = "{fast-resolve: Fast Bayesian radio interferometric imaging}",
      journal = {\aap},
         year = 2024,
        month = oct,
       volume = {690},
          eid = {A387},
        pages = {A387},
          doi = {10.1051/0004-6361/202451107},
archivePrefix = {arXiv},
       eprint = {2406.09144},
 primaryClass = {astro-ph.IM},
       adsurl = {https://ui.adsabs.harvard.edu/abs/2024A&A...690A.387R}
}

@ARTICLE{Roth_23,
       author = {{Roth}, Jakob and {Arras}, Philipp and {Reinecke}, Martin and {Perley}, Richard A. and {Westermann}, R{\"u}diger and {En{\ss}lin}, Torsten A.},
        title = "{Bayesian radio interferometric imaging with direction-dependent calibration}",
      journal = {\aap},
         year = 2023,
        month = oct,
       volume = {678},
          eid = {A177},
        pages = {A177},
          doi = {10.1051/0004-6361/202346851},
archivePrefix = {arXiv},
       eprint = {2305.05489},
 primaryClass = {astro-ph.IM},
       adsurl = {https://ui.adsabs.harvard.edu/abs/2023A&A...678A.177R}
}

@article{Arras_21,
    journal = {Astronomy \& Astrophysics},
    author = {Arras, Philipp and Bester, Hertzog L and Perley, Richard A and Leike, Reimar and Smirnov, Oleg and Westermann, R{\"u}diger and En{\ss}lin, Torsten A},
    year = {2021},
    volume = {646},
    title = {Comparison of classical and Bayesian imaging in radio interferometry-Cygnus A with CLEAN and resolve},
    pages = {A84},
    publisher = {EDP Sciences},
}

@ARTICLE{Arras_25,
       author = {{Arras}, Philipp and {Roth}, Jakob and {Reinecke}, Martin and {Perley}, Richard A. and {Frolov}, Andrei and {Westermann}, R{\"u}diger and {En{\ss}lin}, Torsten A.},
        title = "{Bayesian Imaging of Interferometric Data from Polarized Electromagnetic Signals}",
      journal = {arXiv e-prints},
         year = 2025,
        month = mar,
          eid = {arXiv:2504.00227},
        pages = {arXiv:2504.00227},
          doi = {10.48550/arXiv.2504.00227},
archivePrefix = {arXiv},
       eprint = {2504.00227},
 primaryClass = {astro-ph.IM},
       adsurl = {https://ui.adsabs.harvard.edu/abs/2025arXiv250400227A}
}

@ARTICLE{Honma_2014,
       author = {{Honma}, Mareki and {Akiyama}, Kazunori and {Uemura}, Makoto and {Ikeda}, Shiro},
        title = "{Super-resolution imaging with radio interferometry using sparse modeling}",
      journal = {\pasj},
         year = 2014,
        month = oct,
       volume = {66},
       number = {5},
          eid = {95},
        pages = {95},
          doi = {10.1093/pasj/psu070},
       adsurl = {https://ui.adsabs.harvard.edu/abs/2014PASJ...66...95H}
}

@INPROCEEDINGS{Shepherd_1997_DIFMAP,
       author = {{Shepherd}, M.~C.},
        title = "{Difmap: an Interactive Program for Synthesis Imaging}",
    booktitle = {Astronomical Data Analysis Software and Systems VI},
         year = 1997,
       editor = {{Hunt}, Gareth and {Payne}, Harry},
       series = {Astronomical Society of the Pacific Conference Series},
       volume = {125},
        month = jan,
        pages = {77},
       adsurl = {https://ui.adsabs.harvard.edu/abs/1997ASPC..125...77S}
}

@ARTICLE{JSKIM_2025,
       author = {{Kim}, Jong-Seo and {M{\"u}ller}, Hendrik and {Nikonov}, Aleksei S. and {Lu}, Ru-Sen and {Knollm{\"u}ller}, Jakob and {En{\ss}lin}, Torsten A. and {Wielgus}, Maciek and {Lobanov}, Andrei P.},
        title = "{Imaging a ring-like structure and the extended jet of M87 at 86 GHz}",
      journal = {\aap},
         year = 2025,
        month = apr,
       volume = {696},
          eid = {A169},
        pages = {A169},
          doi = {10.1051/0004-6361/202452038},
       adsurl = {https://ui.adsabs.harvard.edu/abs/2025A&A...696A.169K}
}

@misc{knoll_18,
      title={Encoding prior knowledge in the structure of the likelihood}, 
      author={Jakob Knollmüller and Torsten A. Enßlin},
      year={2018},
      eprint={1812.04403},
      archivePrefix={arXiv},
      primaryClass={stat.ML},
      url={https://arxiv.org/abs/1812.04403}, 
}

@BOOK{Rybicki_1979,
       author = {{Rybicki}, George B. and {Lightman}, Alan P.},
        title = "{Radiative processes in astrophysics}",
         year = 1979,
       adsurl = {https://ui.adsabs.harvard.edu/abs/1979rpa..book.....R}
}

@ARTICLE{Birdi_2018_Polarized_SARA,
       author = {{Birdi}, Jasleen and {Repetti}, Audrey and {Wiaux}, Yves},
        title = "{Sparse interferometric Stokes imaging under the polarization constraint (Polarized SARA)}",
      journal = {\mnras},
         year = 2018,
        month = aug,
       volume = {478},
       number = {4},
        pages = {4442-4463},
          doi = {10.1093/mnras/sty1182},
archivePrefix = {arXiv},
       eprint = {1801.02417},
 primaryClass = {astro-ph.IM},
       adsurl = {https://ui.adsabs.harvard.edu/abs/2018MNRAS.478.4442B}
}

@ARTICLE{Vogt_Ensslin_2005_Bayesian_RM,
       author = {{Vogt}, C. and {En{\ss}lin}, T.~A.},
        title = "{A Bayesian view on Faraday rotation maps   Seeing the magnetic power spectra in galaxy clusters}",
      journal = {\aap},
         year = 2005,
        month = apr,
       volume = {434},
       number = {1},
        pages = {67-76},
          doi = {10.1051/0004-6361:20041839},
archivePrefix = {arXiv},
       eprint = {astro-ph/0501211},
 primaryClass = {astro-ph},
       adsurl = {https://ui.adsabs.harvard.edu/abs/2005A&A...434...67V}
}

@ARTICLE{EHT_2021_M87_polarimetry,
       author = {{Event Horizon Telescope Collaboration} and {Akiyama}, Kazunori and {Algaba}, Juan Carlos and {Alberdi}, Antxon and {Alef}, Walter and {Anantua}, Richard and {Asada}, Keiichi and {Azulay}, Rebecca and {Baczko}, Anne-Kathrin and {Ball}, David and {Balokovi{\'c}}, Mislav and {Barrett}, John and {Benson}, Bradford A. and {Bintley}, Dan and {Blackburn}, Lindy and {Blundell}, Raymond and {Boland}, Wilfred and {Bouman}, Katherine L. and {Bower}, Geoffrey C. and {Boyce}, Hope and {Bremer}, Michael and {Brinkerink}, Christiaan D. and {Brissenden}, Roger and {Britzen}, Silke and {Broderick}, Avery E. and {Broguiere}, Dominique and {Bronzwaer}, Thomas and {Byun}, Do-Young and {Carlstrom}, John E. and {Chael}, Andrew and {Chan}, Chi-kwan and {Chatterjee}, Shami and {Chatterjee}, Koushik and {Chen}, Ming-Tang and {Chen}, Yongjun and {Chesler}, Paul M. and {Cho}, Ilje and {Christian}, Pierre and {Conway}, John E. and {Cordes}, James M. and {Crawford}, Thomas M. and {Crew}, Geoffrey B. and {Cruz-Osorio}, Alejandro and {Cui}, Yuzhu and {Davelaar}, Jordy and {De Laurentis}, Mariafelicia and {Deane}, Roger and {Dempsey}, Jessica and {Desvignes}, Gregory and {Dexter}, Jason and {Doeleman}, Sheperd S. and {Eatough}, Ralph P. and {Falcke}, Heino and {Farah}, Joseph and {Fish}, Vincent L. and {Fomalont}, Ed and {Ford}, H. Alyson and {Fraga-Encinas}, Raquel and {Freeman}, William T. and {Friberg}, Per and {Fromm}, Christian M. and {Fuentes}, Antonio and {Galison}, Peter and {Gammie}, Charles F. and {Garc{\'\i}a}, Roberto and {Gentaz}, Olivier and {Georgiev}, Boris and {Goddi}, Ciriaco and {Gold}, Roman and {G{\'o}mez}, Jos{\'e} L. and {G{\'o}mez-Ruiz}, Arturo I. and {Gu}, Minfeng and {Gurwell}, Mark and {Hada}, Kazuhiro and {Haggard}, Daryl and {Hecht}, Michael H. and {Hesper}, Ronald and {Ho}, Luis C. and {Ho}, Paul and {Honma}, Mareki and {Huang}, Chih-Wei L. and {Huang}, Lei and {Hughes}, David H. and {Ikeda}, Shiro and {Inoue}, Makoto and {Issaoun}, Sara and {James}, David J. and {Jannuzi}, Buell T. and {Janssen}, Michael and {Jeter}, Britton and {Jiang}, Wu and {Jimenez-Rosales}, Alejandra and {Johnson}, Michael D. and {Jorstad}, Svetlana and {Jung}, Taehyun and {Karami}, Mansour and {Karuppusamy}, Ramesh and {Kawashima}, Tomohisa and {Keating}, Garrett K. and {Kettenis}, Mark and {Kim}, Dong-Jin and {Kim}, Jae-Young and {Kim}, Jongsoo and {Kim}, Junhan and {Kino}, Motoki and {Koay}, Jun Yi and {Kofuji}, Yutaro and {Koch}, Patrick M. and {Koyama}, Shoko and {Kramer}, Michael and {Kramer}, Carsten and {Krichbaum}, Thomas P. and {Kuo}, Cheng-Yu and {Lauer}, Tod R. and {Lee}, Sang-Sung and {Levis}, Aviad and {Li}, Yan-Rong and {Li}, Zhiyuan and {Lindqvist}, Michael and {Lico}, Rocco and {Lindahl}, Greg and {Liu}, Jun and {Liu}, Kuo and {Liuzzo}, Elisabetta and {Lo}, Wen-Ping and {Lobanov}, Andrei P. and {Loinard}, Laurent and {Lonsdale}, Colin and {Lu}, Ru-Sen and {MacDonald}, Nicholas R. and {Mao}, Jirong and {Marchili}, Nicola and {Markoff}, Sera and {Marrone}, Daniel P. and {Marscher}, Alan P. and {Mart{\'\i}-Vidal}, Iv{\'a}n and {Matsushita}, Satoki and {Matthews}, Lynn D. and {Medeiros}, Lia and {Menten}, Karl M. and {Mizuno}, Izumi and {Mizuno}, Yosuke and {Moran}, James M. and {Moriyama}, Kotaro and {Moscibrodzka}, Monika and {M{\"u}ller}, Cornelia and {Musoke}, Gibwa and {Mej{\'\i}as}, Alejandro Mus and {Michalik}, Daniel and {Nadolski}, Andrew and {Nagai}, Hiroshi and {Nagar}, Neil M. and {Nakamura}, Masanori and {Narayan}, Ramesh and {Narayanan}, Gopal and {Natarajan}, Iniyan and {Nathanail}, Antonios and {Neilsen}, Joey and {Neri}, Roberto and {Ni}, Chunchong and {Noutsos}, Aristeidis and {Nowak}, Michael A. and {Okino}, Hiroki and {Olivares}, H{\'e}ctor and {Ortiz-Le{\'o}n}, Gisela N. and {Oyama}, Tomoaki and {{\"O}zel}, Feryal and {Palumbo}, Daniel C.~M. and {Park}, Jongho and {Patel}, Nimesh and {Pen}, Ue-Li and {Pesce}, Dominic W. and {Pi{\'e}tu}, Vincent and {Plambeck}, Richard and {PopStefanija}, Aleksandar and {Porth}, Oliver and {P{\"o}tzl}, Felix M. and {Prather}, Ben and {Preciado-L{\'o}pez}, Jorge A. and {Psaltis}, Dimitrios and {Pu}, Hung-Yi and {Ramakrishnan}, Venkatessh and {Rao}, Ramprasad and {Rawlings}, Mark G. and {Raymond}, Alexander W. and {Rezzolla}, Luciano and {Ricarte}, Angelo and {Ripperda}, Bart and {Roelofs}, Freek and {Rogers}, Alan and {Ros}, Eduardo and {Rose}, Mel and {Roshanineshat}, Arash and {Rottmann}, Helge and {Roy}, Alan L. and {Ruszczyk}, Chet and {Rygl}, Kazi L.~J. and {S{\'a}nchez}, Salvador and {S{\'a}nchez-Arguelles}, David and {Sasada}, Mahito},
        title = "{First M87 Event Horizon Telescope Results. VII. Polarization of the Ring}",
      journal = {\apjl},
         year = 2021,
        month = mar,
       volume = {910},
       number = {1},
          eid = {L12},
        pages = {L12},
          doi = {10.3847/2041-8213/abe71d},
archivePrefix = {arXiv},
       eprint = {2105.01169},
 primaryClass = {astro-ph.HE},
       adsurl = {https://ui.adsabs.harvard.edu/abs/2021ApJ...910L..12E}
}

@ARTICLE{EHT_2024_SGRA_polarimetry,
       author = {{Event Horizon Telescope Collaboration} and {Akiyama}, Kazunori and {Alberdi}, Antxon and {Alef}, Walter and {Algaba}, Juan Carlos and {Anantua}, Richard and {Asada}, Keiichi and {Azulay}, Rebecca and {Bach}, Uwe and {Baczko}, Anne-Kathrin and {Ball}, David and {Balokovic}, Mislav and {Bandyopadhyay}, Bidisha and {Barrett}, John and {Baub{\"o}ck}, Michi and {Benson}, Bradford A. and {Bintley}, Dan and {Blackburn}, Lindy and {Blundell}, Raymond and {Bouman}, Katherine L. and {Bower}, Geoffrey C. and {Boyce}, Hope and {Bremer}, Michael and {Brinkerink}, Christiaan D. and {Brissenden}, Roger and {Britzen}, Silke and {Broderick}, Avery E. and {Broguiere}, Dominique and {Bronzwaer}, Thomas and {Bustamante}, Sandra and {Byun}, Do-Young and {Carlstrom}, John E. and {Ceccobello}, Chiara and {Chael}, Andrew and {Chan}, Chi-kwan and {Chang}, Dominic O. and {Chatterjee}, Koushik and {Chatterjee}, Shami and {Chen}, Ming-Tang and {Chen}, Yongjun and {Cheng}, Xiaopeng and {Cho}, Ilje and {Christian}, Pierre and {Conroy}, Nicholas S. and {Conway}, John E. and {Cordes}, James M. and {Crawford}, Thomas M. and {Crew}, Geoffrey B. and {Cruz-Osorio}, Alejandro and {Cui}, Yuzhu and {Dahale}, Rohan and {Davelaar}, Jordy and {De Laurentis}, Mariafelicia and {Deane}, Roger and {Dempsey}, Jessica and {Desvignes}, Gregory and {Dexter}, Jason and {Dhruv}, Vedant and {Dihingia}, Indu K. and {Doeleman}, Sheperd S. and {Dougal}, Sean Taylor and {Dzib}, Sergio A. and {Eatough}, Ralph P. and {Emami}, Razieh and {Falcke}, Heino and {Farah}, Joseph and {Fish}, Vincent L. and {Fomalont}, Edward and {Ford}, H. Alyson and {Foschi}, Marianna and {Fraga-Encinas}, Raquel and {Freeman}, William T. and {Friberg}, Per and {Fromm}, Christian M. and {Fuentes}, Antonio and {Galison}, Peter and {Gammie}, Charles F. and {Garc{\'\i}a}, Roberto and {Gentaz}, Olivier and {Georgiev}, Boris and {Goddi}, Ciriaco and {Gold}, Roman and {G{\'o}mez-Ruiz}, Arturo I. and {G{\'o}mez}, Jos{\'e} L. and {Gu}, Minfeng and {Gurwell}, Mark and {Hada}, Kazuhiro and {Haggard}, Daryl and {Haworth}, Kari and {Hecht}, Michael H. and {Hesper}, Ronald and {Heumann}, Dirk and {Ho}, Luis C. and {Ho}, Paul and {Honma}, Mareki and {Huang}, Chih-Wei L. and {Huang}, Lei and {Hughes}, David H. and {Ikeda}, Shiro and {Impellizzeri}, C.~M. Violette and {Inoue}, Makoto and {Issaoun}, Sara and {James}, David J. and {Jannuzi}, Buell T. and {Janssen}, Michael and {Jeter}, Britton and {Jiang}, Wu and {Jim{\'e}nez-Rosales}, Alejandra and {Johnson}, Michael D. and {Jorstad}, Svetlana and {Joshi}, Abhishek V. and {Jung}, Taehyun and {Karami}, Mansour and {Karuppusamy}, Ramesh and {Kawashima}, Tomohisa and {Keating}, Garrett K. and {Kettenis}, Mark and {Kim}, Dong-Jin and {Kim}, Jae-Young and {Kim}, Jongsoo and {Kim}, Junhan and {Kino}, Motoki and {Koay}, Jun Yi and {Kocherlakota}, Prashant and {Kofuji}, Yutaro and {Koch}, Patrick M. and {Koyama}, Shoko and {Kramer}, Carsten and {Kramer}, Joana A. and {Kramer}, Michael and {Krichbaum}, Thomas P. and {Kuo}, Cheng-Yu and {La Bella}, Noemi and {Lauer}, Tod R. and {Lee}, Daeyoung and {Lee}, Sang-Sung and {Leung}, Po Kin and {Levis}, Aviad and {Li}, Zhiyuan and {Lico}, Rocco and {Lindahl}, Greg and {Lindqvist}, Michael and {Lisakov}, Mikhail and {Liu}, Jun and {Liu}, Kuo and {Liuzzo}, Elisabetta and {Lo}, Wen-Ping and {Lobanov}, Andrei P. and {Loinard}, Laurent and {Lonsdale}, Colin J. and {Lowitz}, Amy E. and {Lu}, Ru-Sen and {MacDonald}, Nicholas R. and {Mao}, Jirong and {Marchili}, Nicola and {Markoff}, Sera and {Marrone}, Daniel P. and {Marscher}, Alan P. and {Mart{\'\i}-Vidal}, Iv{\'a}n and {Matsushita}, Satoki and {Matthews}, Lynn D. and {Medeiros}, Lia and {Menten}, Karl M. and {Michalik}, Daniel and {Mizuno}, Izumi and {Mizuno}, Yosuke and {Moran}, James M. and {Moriyama}, Kotaro and {Moscibrodzka}, Monika and {Mulaudzi}, Wanga and {M{\"u}ller}, Cornelia and {M{\"u}ller}, Hendrik and {Mus}, Alejandro and {Musoke}, Gibwa and {Myserlis}, Ioannis and {Nadolski}, Andrew and {Nagai}, Hiroshi and {Nagar}, Neil M. and {Nakamura}, Masanori and {Narayanan}, Gopal and {Natarajan}, Iniyan and {Nathanail}, Antonios and {Fuentes}, Santiago Navarro and {Neilsen}, Joey and {Neri}, Roberto and {Ni}, Chunchong and {Noutsos}, Aristeidis and {Nowak}, Michael A. and {Oh}, Junghwan and {Okino}, Hiroki and {Olivares}, H{\`e}ctor and {Ortiz-Le{\'o}n}, Gisela N. and {Oyama}, Tomoaki and {{\"O}zel}, Feryal and {Palumbo}, Daniel C.~M. and {Paraschos}, Georgios Filippos and {Park}, Jongho and {Parsons}, Harriet and {Patel}, Nimesh and {Pen}, Ue-Li},
        title = "{First Sagittarius A* Event Horizon Telescope Results. VII. Polarization of the Ring}",
      journal = {\apjl},
         year = 2024,
        month = apr,
       volume = {964},
       number = {2},
          eid = {L25},
        pages = {L25},
          doi = {10.3847/2041-8213/ad2df0},
       adsurl = {https://ui.adsabs.harvard.edu/abs/2024ApJ...964L..25E}
}

@ARTICLE{EHT_2025_M87_2021,
       author = {{Event Horizon Telescope Collaboration}},
        title = "{Horizon-scale variability of M87* from 2017--2021 EHT observations}",
      journal = {arXiv e-prints},
         year = 2025,
        month = sep,
          eid = {arXiv:2509.24593},
        pages = {arXiv:2509.24593},
          doi = {10.48550/arXiv.2509.24593},
archivePrefix = {arXiv},
       eprint = {2509.24593},
 primaryClass = {astro-ph.HE},
       adsurl = {https://ui.adsabs.harvard.edu/abs/2025arXiv250924593T}
}

@ARTICLE{2020_Birdi_polcal_SARA,
       author = {{Birdi}, Jasleen and {Repetti}, Audrey and {Wiaux}, Yves},
        title = "{Polca SARA - full polarization, direction-dependent calibration, and sparse imaging for radio interferometry}",
      journal = {\mnras},
         year = 2020,
        month = mar,
       volume = {492},
       number = {3},
        pages = {3509-3528},
          doi = {10.1093/mnras/stz3555},
archivePrefix = {arXiv},
       eprint = {1904.00663},
 primaryClass = {astro-ph.IM},
       adsurl = {https://ui.adsabs.harvard.edu/abs/2020MNRAS.492.3509B}
}

@ARTICLE{2021_Pesce_DMC,
       author = {{Pesce}, Dominic W.},
        title = "{A D-term Modeling Code (DMC) for Simultaneous Calibration and Full-Stokes Imaging of Very Long Baseline Interferometric Data}",
      journal = {\aj},
         year = 2021,
        month = apr,
       volume = {161},
       number = {4},
          eid = {178},
        pages = {178},
          doi = {10.3847/1538-3881/abe3f8},
archivePrefix = {arXiv},
       eprint = {2102.03328},
 primaryClass = {astro-ph.IM},
       adsurl = {https://ui.adsabs.harvard.edu/abs/2021AJ....161..178P}
}

@ARTICLE{Knollmueller_2019_MGVI,
       author = {{Knollm{\"u}ller}, Jakob and {En{\ss}lin}, Torsten A.},
        title = "{Metric Gaussian Variational Inference}",
      journal = {arXiv e-prints},
         year = 2019,
        month = jan,
          eid = {arXiv:1901.11033},
        pages = {arXiv:1901.11033},
          doi = {10.48550/arXiv.1901.11033},
archivePrefix = {arXiv},
       eprint = {1901.11033},
 primaryClass = {stat.ML},
       adsurl = {https://ui.adsabs.harvard.edu/abs/2019arXiv190111033K}
}

@ARTICLE{Frank_2021_geoVI,
       author = {{Frank}, Philipp and {Leike}, Reimar and {En{\ss}lin}, Torsten A.},
        title = "{Geometric Variational Inference}",
      journal = {Entropy},
         year = 2021,
        month = jul,
       volume = {23},
       number = {7},
          eid = {853},
        pages = {853},
          doi = {10.3390/e23070853},
archivePrefix = {arXiv},
       eprint = {2105.10470},
 primaryClass = {stat.ME},
       adsurl = {https://ui.adsabs.harvard.edu/abs/2021Entrp..23..853F}
}

@ARTICLE{EHT_2019_M87_paper1,
       author = {{Event Horizon Telescope Collaboration} and {Akiyama}, Kazunori and {Alberdi}, Antxon and {Alef}, Walter and {Asada}, Keiichi and {Azulay}, Rebecca and {Baczko}, Anne-Kathrin and {Ball}, David and {Balokovi{\'c}}, Mislav and {Barrett}, John and {Bintley}, Dan and {Blackburn}, Lindy and {Boland}, Wilfred and {Bouman}, Katherine L. and {Bower}, Geoffrey C. and {Bremer}, Michael and {Brinkerink}, Christiaan D. and {Brissenden}, Roger and {Britzen}, Silke and {Broderick}, Avery E. and {Broguiere}, Dominique and {Bronzwaer}, Thomas and {Byun}, Do-Young and {Carlstrom}, John E. and {Chael}, Andrew and {Chan}, Chi-kwan and {Chatterjee}, Shami and {Chatterjee}, Koushik and {Chen}, Ming-Tang and {Chen}, Yongjun and {Cho}, Ilje and {Christian}, Pierre and {Conway}, John E. and {Cordes}, James M. and {Crew}, Geoffrey B. and {Cui}, Yuzhu and {Davelaar}, Jordy and {De Laurentis}, Mariafelicia and {Deane}, Roger and {Dempsey}, Jessica and {Desvignes}, Gregory and {Dexter}, Jason and {Doeleman}, Sheperd S. and {Eatough}, Ralph P. and {Falcke}, Heino and {Fish}, Vincent L. and {Fomalont}, Ed and {Fraga-Encinas}, Raquel and {Freeman}, William T. and {Friberg}, Per and {Fromm}, Christian M. and {G{\'o}mez}, Jos{\'e} L. and {Galison}, Peter and {Gammie}, Charles F. and {Garc{\'\i}a}, Roberto and {Gentaz}, Olivier and {Georgiev}, Boris and {Goddi}, Ciriaco and {Gold}, Roman and {Gu}, Minfeng and {Gurwell}, Mark and {Hada}, Kazuhiro and {Hecht}, Michael H. and {Hesper}, Ronald and {Ho}, Luis C. and {Ho}, Paul and {Honma}, Mareki and {Huang}, Chih-Wei L. and {Huang}, Lei and {Hughes}, David H. and {Ikeda}, Shiro and {Inoue}, Makoto and {Issaoun}, Sara and {James}, David J. and {Jannuzi}, Buell T. and {Janssen}, Michael and {Jeter}, Britton and {Jiang}, Wu and {Johnson}, Michael D. and {Jorstad}, Svetlana and {Jung}, Taehyun and {Karami}, Mansour and {Karuppusamy}, Ramesh and {Kawashima}, Tomohisa and {Keating}, Garrett K. and {Kettenis}, Mark and {Kim}, Jae-Young and {Kim}, Junhan and {Kim}, Jongsoo and {Kino}, Motoki and {Koay}, Jun Yi and {Koch}, Patrick M. and {Koyama}, Shoko and {Kramer}, Michael and {Kramer}, Carsten and {Krichbaum}, Thomas P. and {Kuo}, Cheng-Yu and {Lauer}, Tod R. and {Lee}, Sang-Sung and {Li}, Yan-Rong and {Li}, Zhiyuan and {Lindqvist}, Michael and {Liu}, Kuo and {Liuzzo}, Elisabetta and {Lo}, Wen-Ping and {Lobanov}, Andrei P. and {Loinard}, Laurent and {Lonsdale}, Colin and {Lu}, Ru-Sen and {MacDonald}, Nicholas R. and {Mao}, Jirong and {Markoff}, Sera and {Marrone}, Daniel P. and {Marscher}, Alan P. and {Mart{\'\i}-Vidal}, Iv{\'a}n and {Matsushita}, Satoki and {Matthews}, Lynn D. and {Medeiros}, Lia and {Menten}, Karl M. and {Mizuno}, Yosuke and {Mizuno}, Izumi and {Moran}, James M. and {Moriyama}, Kotaro and {Moscibrodzka}, Monika and {M{\"u}ller}, Cornelia and {Nagai}, Hiroshi and {Nagar}, Neil M. and {Nakamura}, Masanori and {Narayan}, Ramesh and {Narayanan}, Gopal and {Natarajan}, Iniyan and {Neri}, Roberto and {Ni}, Chunchong and {Noutsos}, Aristeidis and {Okino}, Hiroki and {Olivares}, H{\'e}ctor and {Ortiz-Le{\'o}n}, Gisela N. and {Oyama}, Tomoaki and {{\"O}zel}, Feryal and {Palumbo}, Daniel C.~M. and {Patel}, Nimesh and {Pen}, Ue-Li and {Pesce}, Dominic W. and {Pi{\'e}tu}, Vincent and {Plambeck}, Richard and {PopStefanija}, Aleksandar and {Porth}, Oliver and {Prather}, Ben and {Preciado-L{\'o}pez}, Jorge A. and {Psaltis}, Dimitrios and {Pu}, Hung-Yi and {Ramakrishnan}, Venkatessh and {Rao}, Ramprasad and {Rawlings}, Mark G. and {Raymond}, Alexander W. and {Rezzolla}, Luciano and {Ripperda}, Bart and {Roelofs}, Freek and {Rogers}, Alan and {Ros}, Eduardo and {Rose}, Mel and {Roshanineshat}, Arash and {Rottmann}, Helge and {Roy}, Alan L. and {Ruszczyk}, Chet and {Ryan}, Benjamin R. and {Rygl}, Kazi L.~J. and {S{\'a}nchez}, Salvador and {S{\'a}nchez-Arguelles}, David and {Sasada}, Mahito and {Savolainen}, Tuomas and {Schloerb}, F. Peter and {Schuster}, Karl-Friedrich and {Shao}, Lijing and {Shen}, Zhiqiang and {Small}, Des and {Sohn}, Bong Won and {SooHoo}, Jason and {Tazaki}, Fumie and {Tiede}, Paul and {Tilanus}, Remo P.~J. and {Titus}, Michael and {Toma}, Kenji and {Torne}, Pablo and {Trent}, Tyler and {Trippe}, Sascha and {Tsuda}, Shuichiro and {van Bemmel}, Ilse and {van Langevelde}, Huib Jan and {van Rossum}, Daniel R. and {Wagner}, Jan and {Wardle}, John and {Weintroub}, Jonathan and {Wex}, Norbert and {Wharton}, Robert and {Wielgus}, Maciek and {Wong}, George N. and {Wu}, Qingwen and {Young}, Ken and {Young}, Andr{\'e}},
        title = "{First M87 Event Horizon Telescope Results. I. The Shadow of the Supermassive Black Hole}",
      journal = {\apjl},
         year = 2019,
        month = apr,
       volume = {875},
       number = {1},
          eid = {L1},
        pages = {L1},
          doi = {10.3847/2041-8213/ab0ec7},
archivePrefix = {arXiv},
       eprint = {1906.11238},
 primaryClass = {astro-ph.GA},
       adsurl = {https://ui.adsabs.harvard.edu/abs/2019ApJ...875L...1E}
}

@ARTICLE{EHT_2024_345GHz,
       author = {{Raymond}, Alexander W. and {Doeleman}, Sheperd S. and {Asada}, Keiichi and {Blackburn}, Lindy and {Bower}, Geoffrey C. and {Bremer}, Michael and {Broguiere}, Dominique and {Chen}, Ming-Tang and {Crew}, Geoffrey B. and {Dornbusch}, Sven and {Fish}, Vincent L. and {Garc{\'\i}a}, Roberto and {Gentaz}, Olivier and {Goddi}, Ciriaco and {Han}, Chih-Chiang and {Hecht}, Michael H. and {Huang}, Yau-De and {Janssen}, Michael and {Keating}, Garrett K. and {Koay}, Jun Yi and {Krichbaum}, Thomas P. and {Lo}, Wen-Ping and {Matsushita}, Satoki and {Matthews}, Lynn D. and {Moran}, James M. and {Norton}, Timothy J. and {Patel}, Nimesh and {Pesce}, Dominic W. and {Ramakrishnan}, Venkatessh and {Rottmann}, Helge and {Roy}, Alan L. and {S{\'a}nchez}, Salvador and {Tilanus}, Remo P.~J. and {Titus}, Michael and {Torne}, Pablo and {Wagner}, Jan and {Weintroub}, Jonathan and {Wielgus}, Maciek and {Young}, Andr{\'e} and {Akiyama}, Kazunori and {Albentosa-Ru{\'\i}z}, Ezequiel and {Alberdi}, Antxon and {Alef}, Walter and {Algaba}, Juan Carlos and {Anantua}, Richard and {Azulay}, Rebecca and {Bach}, Uwe and {Baczko}, Anne-Kathrin and {Ball}, David and {Balokovic}, Mislav and {Bandyopadhyay}, Bidisha and {Barrett}, John and {Baub{\"o}ck}, Michi and {Benson}, Bradford A. and {Bintley}, Dan and {Blundell}, Raymond and {Bouman}, Katherine L. and {Boyce}, Hope and {Brissenden}, Roger and {Britzen}, Silke and {Broderick}, Avery E. and {Bronzwaer}, Thomas and {Bustamante}, Sandra and {Carlstrom}, John E. and {Chael}, Andrew and {Chan}, Chi-kwan and {Chang}, Dominic O. and {Chatterjee}, Koushik and {Chatterjee}, Shami and {Chen}, Yongjun and {Cheng}, Xiaopeng and {Cho}, Ilje and {Christian}, Pierre and {Conroy}, Nicholas S. and {Conway}, John E. and {Crawford}, Thomas M. and {Cruz-Osorio}, Alejandro and {Cui}, Yuzhu and {Dahale}, Rohan and {Davelaar}, Jordy and {De Laurentis}, Mariafelicia and {Deane}, Roger and {Dempsey}, Jessica and {Desvignes}, Gregory and {Dexter}, Jason and {Dhruv}, Vedant and {Dihingia}, Indu K. and {Dzib}, Sergio A. and {Eatough}, Ralph P. and {Emami}, Razieh and {Falcke}, Heino and {Farah}, Joseph and {Fomalont}, Edward and {Fontana}, Anne-Laure and {Ford}, H. Alyson and {Foschi}, Marianna and {Fraga-Encinas}, Raquel and {Freeman}, William T. and {Friberg}, Per and {Fromm}, Christian M. and {Fuentes}, Antonio and {Galison}, Peter and {Gammie}, Charles F. and {Georgiev}, Boris and {Gold}, Roman and {G{\'o}mez-Ruiz}, Arturo I. and {G{\'o}mez}, Jos{\'e} L. and {Gu}, Minfeng and {Gurwell}, Mark and {Hada}, Kazuhiro and {Haggard}, Daryl and {Hesper}, Ronald and {Heumann}, Dirk and {Ho}, Luis C. and {Ho}, Paul and {Honma}, Mareki and {Huang}, Chih-Wei L. and {Huang}, Lei and {Hughes}, David H. and {Ikeda}, Shiro and {Impellizzeri}, C.~M. Violette and {Inoue}, Makoto and {Issaoun}, Sara and {James}, David J. and {Jannuzi}, Buell T. and {Jeter}, Britton and {Jiang}, Wu and {Jim{\'e}nez-Rosales}, Alejandra and {Johnson}, Michael D. and {Jorstad}, Svetlana and {Jones}, Adam C. and {Joshi}, Abhishek V. and {Jung}, Taehyun and {Karuppusamy}, Ramesh and {Kawashima}, Tomohisa and {Kettenis}, Mark and {Kim}, Dong-Jin and {Kim}, Jae-Young and {Kim}, Jongsoo and {Kim}, Junhan and {Kino}, Motoki and {Kocherlakota}, Prashant and {Kofuji}, Yutaro and {Koch}, Patrick M. and {Koyama}, Shoko and {Kramer}, Carsten and {Kramer}, Joana A. and {Kramer}, Michael and {Kubo}, Derek and {Kuo}, Cheng-Yu and {La Bella}, Noemi and {Lee}, Sang-Sung and {Levis}, Aviad and {Li}, Zhiyuan and {Lico}, Rocco and {Lindahl}, Greg and {Lindqvist}, Michael and {Lisakov}, Mikhail and {Liu}, Jun and {Liu}, Kuo and {Liuzzo}, Elisabetta and {Lobanov}, Andrei P. and {Loinard}, Laurent and {Lonsdale}, Colin J. and {Lowitz}, Amy E. and {Lu}, Ru-Sen and {MacDonald}, Nicholas R. and {Mahieu}, Sylvain and {Maier}, Doris and {Mao}, Jirong and {Marchili}, Nicola and {Markoff}, Sera and {Marrone}, Daniel P. and {Marscher}, Alan P. and {Mart{\'\i}-Vidal}, Iv{\'a}n and {Medeiros}, Lia and {Menten}, Karl M. and {Mizuno}, Izumi and {Mizuno}, Yosuke and {Montgomery}, Joshua and {Moriyama}, Kotaro and {Moscibrodzka}, Monika and {Mulaudzi}, Wanga and {M{\"u}ller}, Cornelia and {M{\"u}ller}, Hendrik and {Mus}, Alejandro and {Musoke}, Gibwa and {Myserlis}, Ioannis and {Nagai}, Hiroshi and {Nagar}, Neil M. and {Nakamura}, Masanori and {Narayanan}, Gopal and {Natarajan}, Iniyan and {Nathanail}, Antonios and {Fuentes}, Santiago Navarro and {Neilsen}, Joey and {Ni}, Chunchong and {Nowak}, Michael A. and {Oh}, Junghwan and {Okino}, Hiroki},
        title = "{First Very Long Baseline Interferometry Detections at 870 {\ensuremath{\mu}}m}",
      journal = {\aj},
         year = 2024,
        month = sep,
       volume = {168},
       number = {3},
          eid = {130},
        pages = {130},
          doi = {10.3847/1538-3881/ad5bdb},
archivePrefix = {arXiv},
       eprint = {2410.07453},
 primaryClass = {astro-ph.IM},
       adsurl = {https://ui.adsabs.harvard.edu/abs/2024AJ....168..130R}
}

@ARTICLE{Popkov_2021,
       author = {{Popkov}, A.~V. and {Kovalev}, Y.~Y. and {Petrov}, L.~Y. and {Kovalev}, Yu. A.},
        title = "{Parsec-scale Properties of Steep- and Flat-spectrum Extragalactic Radio Sources from a VLBA Survey of a Complete North Polar Cap Sample}",
      journal = {\aj},
         year = 2021,
        month = feb,
       volume = {161},
       number = {2},
          eid = {88},
        pages = {88},
          doi = {10.3847/1538-3881/abd18c},
archivePrefix = {arXiv},
       eprint = {2008.06803},
 primaryClass = {astro-ph.GA},
       adsurl = {https://ui.adsabs.harvard.edu/abs/2021AJ....161...88P}
}

@ARTICLE{Martividal_2008,
       author = {{Mart{\'\i}-Vidal}, I. and {Marcaide}, J.~M.},
        title = "{Spurious source generation in mapping from noisy phase-self-calibrated data}",
      journal = {\aap},
         year = 2008,
        month = mar,
       volume = {480},
       number = {1},
        pages = {289-295},
          doi = {10.1051/0004-6361:20078690},
archivePrefix = {arXiv},
       eprint = {0801.1272},
 primaryClass = {astro-ph},
       adsurl = {https://ui.adsabs.harvard.edu/abs/2008A&A...480..289M}
}

@ARTICLE{Cornwell_2008,
       author = {{Cornwell}, T.~J.},
        title = "{Multiscale CLEAN Deconvolution of Radio Synthesis Images}",
      journal = {IEEE Journal of Selected Topics in Signal Processing},
         year = 2008,
        month = nov,
       volume = {2},
       number = {5},
        pages = {793-801},
          doi = {10.1109/JSTSP.2008.2006388},
       adsurl = {https://ui.adsabs.harvard.edu/abs/2008ISTSP...2..793C}
}

@ARTICLE{Janssen_2019_rPICARD,
       author = {{Janssen}, M. and {Goddi}, C. and {van Bemmel}, I.~M. and {Kettenis}, M. and {Small}, D. and {Liuzzo}, E. and {Rygl}, K. and {Mart{\'\i}-Vidal}, I. and {Blackburn}, L. and {Wielgus}, M. and {Falcke}, H.},
        title = "{rPICARD: A CASA-based calibration pipeline for VLBI data. Calibration and imaging of 7 mm VLBA observations of the AGN jet in M 87}",
      journal = {\aap},
         year = 2019,
        month = jun,
       volume = {626},
          eid = {A75},
        pages = {A75},
          doi = {10.1051/0004-6361/201935181},
archivePrefix = {arXiv},
       eprint = {1905.01905},
 primaryClass = {astro-ph.IM},
       adsurl = {https://ui.adsabs.harvard.edu/abs/2019A&A...626A..75J}
}

@ARTICLE{CASA_2022,
       author = {{CASA Team} and {Bean}, Ben and {Bhatnagar}, Sanjay and {Castro}, Sandra and {Donovan Meyer}, Jennifer and {Emonts}, Bjorn and {Garcia}, Enrique and {Garwood}, Robert and {Golap}, Kumar and {Gonzalez Villalba}, Justo and {Harris}, Pamela and {Hayashi}, Yohei and {Hoskins}, Josh and {Hsieh}, Mingyu and {Jagannathan}, Preshanth and {Kawasaki}, Wataru and {Keimpema}, Aard and {Kettenis}, Mark and {Lopez}, Jorge and {Marvil}, Joshua and {Masters}, Joseph and {McNichols}, Andrew and {Mehringer}, David and {Miel}, Renaud and {Moellenbrock}, George and {Montesino}, Federico and {Nakazato}, Takeshi and {Ott}, Juergen and {Petry}, Dirk and {Pokorny}, Martin and {Raba}, Ryan and {Rau}, Urvashi and {Schiebel}, Darrell and {Schweighart}, Neal and {Sekhar}, Srikrishna and {Shimada}, Kazuhiko and {Small}, Des and {Steeb}, Jan-Willem and {Sugimoto}, Kanako and {Suoranta}, Ville and {Tsutsumi}, Takahiro and {van Bemmel}, Ilse M. and {Verkouter}, Marjolein and {Wells}, Akeem and {Xiong}, Wei and {Szomoru}, Arpad and {Griffith}, Morgan and {Glendenning}, Brian and {Kern}, Jeff},
        title = "{CASA, the Common Astronomy Software Applications for Radio Astronomy}",
      journal = {\pasp},
         year = 2022,
        month = nov,
       volume = {134},
       number = {1041},
          eid = {114501},
        pages = {114501},
          doi = {10.1088/1538-3873/ac9642},
archivePrefix = {arXiv},
       eprint = {2210.02276},
 primaryClass = {astro-ph.IM},
       adsurl = {https://ui.adsabs.harvard.edu/abs/2022PASP..134k4501C}
}

@ARTICLE{Sillanpaa_1988,
       author = {{Sillanpaa}, A. and {Haarala}, S. and {Valtonen}, M.~J. and {Sundelius}, B. and {Byrd}, G.~G.},
        title = "{OJ 287: Binary Pair of Supermassive Black Holes}",
      journal = {\apj},
         year = 1988,
        month = feb,
       volume = {325},
        pages = {628},
          doi = {10.1086/166033},
       adsurl = {https://ui.adsabs.harvard.edu/abs/1988ApJ...325..628S}
}

@ARTICLE{Roberts_1994,
       author = {{Roberts}, D.~H. and {Wardle}, J.~F.~C. and {Brown}, L.~F.},
        title = "{Linear Polarization Radio Imaging at Milliarcsecond Resolution}",
      journal = {\apj},
         year = 1994,
        month = jun,
       volume = {427},
        pages = {718},
          doi = {10.1086/174180},
       adsurl = {https://ui.adsabs.harvard.edu/abs/1994ApJ...427..718R}
}

@ARTICLE{Lister_2021,
       author = {{Lister}, M.~L. and {Homan}, D.~C. and {Kellermann}, K.~I. and {Kovalev}, Y.~Y. and {Pushkarev}, A.~B. and {Ros}, E. and {Savolainen}, T.},
        title = "{Monitoring Of Jets in Active Galactic Nuclei with VLBA Experiments. XVIII. Kinematics and Inner Jet Evolution of Bright Radio-loud Active Galaxies}",
      journal = {\apj},
         year = 2021,
        month = dec,
       volume = {923},
       number = {1},
          eid = {30},
        pages = {30},
          doi = {10.3847/1538-4357/ac230f},
archivePrefix = {arXiv},
       eprint = {2108.13358},
 primaryClass = {astro-ph.HE},
       adsurl = {https://ui.adsabs.harvard.edu/abs/2021ApJ...923...30L}
}

@book{Wiener_1949, place={Cambridge, Mass}, title={Extrapolation, interpolation, and smoothing of stationary time series, with engineering applications}, publisher={MIT Press}, author={Wiener, Herbert}, year={1949}}

@article{Khinchin_1934,
author = {Khinchin, A.},
journal = {Mathematische Annalen},
pages = {604-615},
title = {Korrelationstheorie der stationären stochastischen Prozesse},
url = {http://eudml.org/doc/159698},
volume = {109},
year = {1934},
}
\appendix

\section{Comparison of VLBA 3C273 \texttt{resolve} reconstructions}

\begin{figure}
    \centering
    \includegraphics[width=9cm]{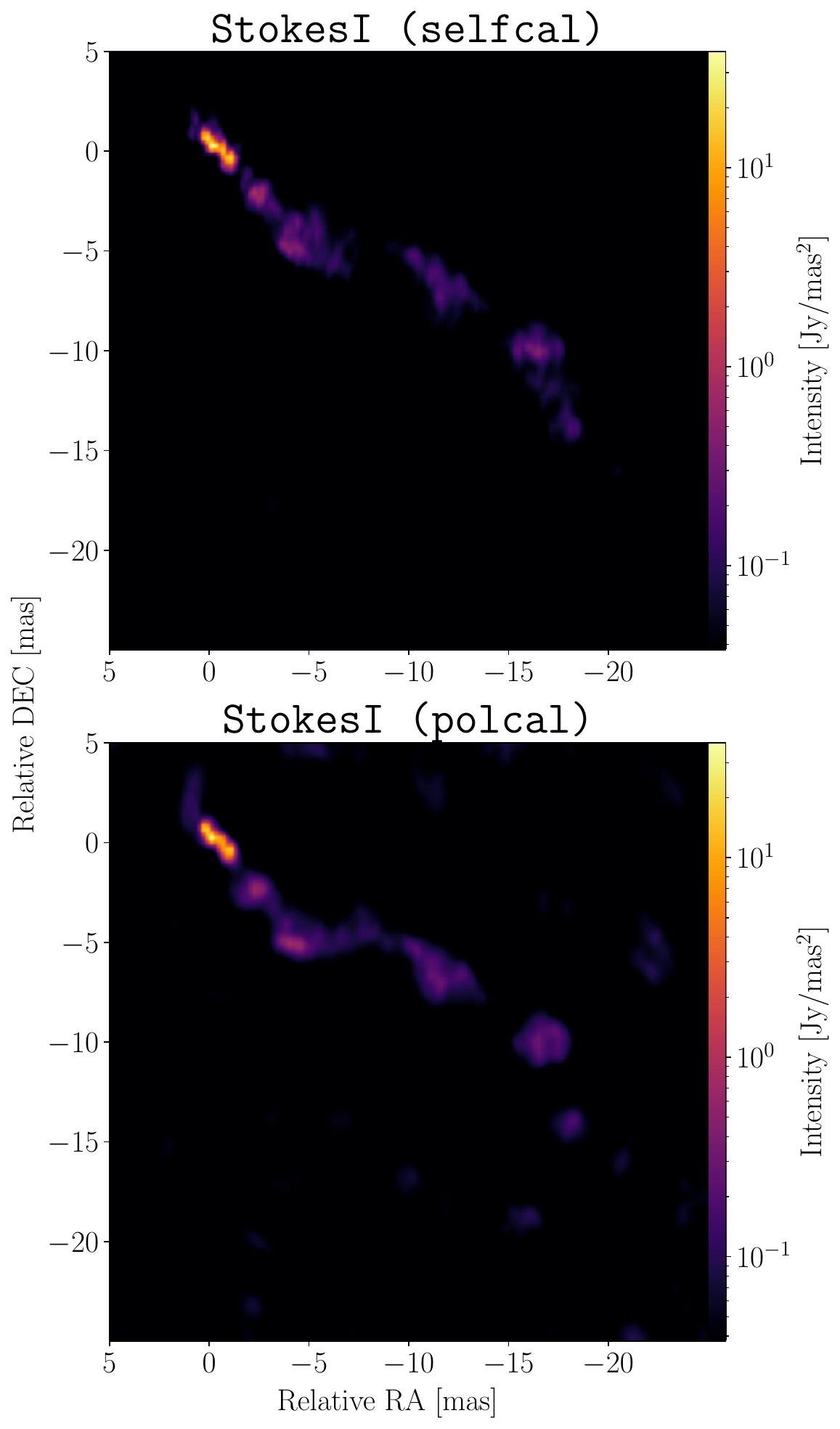}
    \caption{Comparison between the 3C273 VLBA \texttt{resolve} posterior mean Stokes I reconstructions at 15 GHz using self-calibration and polarization calibration. The colorbar starts from 0.1$\%$ of the peak total intensity of \texttt{resolve} image using the polarization calibration.}\label{3C273_StokesI_comparison}
\end{figure} 

\autoref{3C273_StokesI_comparison} represents the comparison between VLBA 3C273 Stokes I \texttt{resolve} posterior mean reconstructions using Bayesian self-calibration and imaging method (left figure) and Bayesian polarization calibration and imaging method (right figure). We note that VLBI data can have higher LR, RL amplitudes than RR, LL amplitudes in some baselines. Even if we do not enforce the polarization constraint in the visibility domain, encoding the polarization constraint in the image domain might generate spurious Stokes I emission. To validate this issue, we performed Bayesian self-calibration and imaging method \citep{JSKim_2024} using VLBA 3C273 multi-IF data. Two images look consistent and the Pearson correlation coefficient between two images is 0.96. In conclusion, the polarization constraint does not produce spurious Stokes I emission for the 3C273 data. However, the Stokes I image from the polarization calibration and imaging method (right panel) has slightly more artifacts outside of the emission region. It might result from the sparse UV-coverage (The on-source time is around 1 hour), which complicates the solution of the degenerate inverse problem.

\section{Hyperparameter setup for the image and gain prior model}\label{Appendix_hyperparameter}

In the Gaussian process prior model in \autoref{eq:Gaussian_process}, the offset mean represents the logarithmic mean flux of the field in the unit of $\text{Jy/str}$. The zero mode variance mean and zero mode variance std correspond to the standard deviation of the offset mean and standard deviation of the standard deviation of the offset mean respectively. The fluctuation activates the fluctuation of the field, related to the dynamic range of the image. The flexibility and asperity are related to stochastic processes, such as the integrated Wiener process and Wiener process in the model. The flexibility parameter ensures flexibility of the power spectrum shape and asperity activates certain amplitude mode in the power spectrum, which can generate periodic patterns in the field. Average slope represents the slope of the power spectrum. As an example, a steep power spectrum corresponds to a smooth image since high Fourier modes are suppressed. The exact mathematical definition of the generative Gaussian process model can be found in \citet[Sec.~3.4]{Arras_21}.

\autoref{table:model_parameters_polsky_3C273} and \autoref{table:model_parameters_cal_3C273} show the hyperparameter setups for the VLBA 3C273 at 15 GHz \texttt{resolve} image prior and calibration prior respectively. The hyperparameter setups for the GMVA+ALMA OJ287 at 86 GHz \texttt{resolve} image prior and calibration prior are in \autoref{table:model_parameters_polsky_OJ287} and \autoref{table:model_parameters_cal_OJ287} correspondingly.

\begin{table*}[h]
 	\caption{Model parameters for the \texttt{resolve} 3C273 polarization imaging priors $s$, $q$, $u$, and $v$.}
    \centering
	\begin{tabularx}{14.5cm}{lrrrrrr}
		\Xhline{2\arrayrulewidth}
		$\:$               & $ \: \: \: \: \: \: \: \: s$ mean & \: \: $s$ std & \: \: $q,u$ mean & \: \: $q,u$ std & \: \: $v$ mean & \: \: $v$ std\\
		\hline
		Offset 	           & 35.0 & --- & 0.0 & --- & 0.0 & ---\\
		Zero mode variance & 1.0 & 0.1 & 0.01 & 0.01 & 0.001 & 0.001\\
		Fluctuations       & 3.0 & 1.0 & 0.05 & 0.05 & 0.005 & 0.005\\
		Flexibility        & 1.2 & 0.4 & 0.1 & 0.1 & 0.1 & 0.1 \\
		Asperity           & --- & --- & --- & --- & --- & --- \\
		Average slope      & -3.0 & 1.0 & -3.0 & 1.0 & -3.0 & 1.0  \\
		\Xhline{2\arrayrulewidth}
	\end{tabularx}
	\label{table:model_parameters_polsky_3C273}
\end{table*}

\begin{table*}[h]
    \caption{Model parameters for the \texttt{resolve} 3C273 log-amplitude gain prior $\lambda$, phase gain prior $\phi$, log-amplitude D-term prior $a$, and phase D-term prior $b$.}
	\centering
	\begin{tabular}{lrrrrrrr}
		\Xhline{2\arrayrulewidth}
		$\:$               & $\lambda$ mean & $\lambda$ std & $\phi$ mean  & $\phi$ std & & $a$ & $b$\\
		\hline
		Offset 	           & 0.0 & --- & 0.0 & --- & Mean     & -2.5 & 0.0\\
		Zero mode variance & 0.1 & 0.01 & 0.2 & 0.1 & Std     & 1.0 & 3.0  \\
		Fluctuations       & 0.2 & 0.1 & 0.2 & 0.1 \\
		Flexibility        & 0.5 & 0.2 & 0.5 & 0.2 \\
		Asperity           & --- & --- & --- & --- \\
		Average slope      & -3.0 & 1.0 & -3.0 & 1.0 \\
		\Xhline{2\arrayrulewidth}
	\end{tabular}\label{table:model_parameters_cal_3C273}
\end{table*}

\begin{table*}[h]
 	\caption{Model parameters for the \texttt{resolve} OJ287 polarization imaging priors $s$, $q$, $u$, and $v$.}
    \centering
	\begin{tabularx}{14.5cm}{lrrrrrr}
		\Xhline{2\arrayrulewidth}
		$\:$               & $ \: \: \: \: \: \: \: \: s$ mean & \: \: $s$ std & \: \: $q,u$ mean & \: \: $q,u$ std & \: \: $v$ mean & \: \: $v$ std\\
		\hline
		Offset 	           & 38.0 & --- & 0.0 & --- & 0.0 & ---\\
		Zero mode variance & 1.0 & 0.1 & 0.01 & 0.01 & 0.001 & 0.001\\
		Fluctuations       & 3.0 & 1.0 & 0.05 & 0.05 & 0.005 & 0.005\\
		Flexibility        & 1.2 & 0.4 & 0.1 & 0.1 & 0.1 & 0.1 \\
		Asperity           & --- & --- & --- & --- & --- & --- \\
		Average slope      & -3.0 & 0.5 & -3.0 & 0.5 & -3.0 & 0.5  \\
		\Xhline{2\arrayrulewidth}
	\end{tabularx}
	\label{table:model_parameters_polsky_OJ287}
\end{table*}

\begin{table}[h]
    \caption{Model parameters for the \texttt{resolve} OJ287 log-amplitude gain prior $\lambda$, phase gain prior $\phi$, log-amplitude D-term prior $a$, and phase D-term prior $b$.}
	\centering
	\begin{tabular}{lrrrrrr}
		\Xhline{2\arrayrulewidth}
		$\:$               & $\lambda$ mean & $\lambda$ std &  & $\phi$ & $a$ & $b$\\
		\hline
		Offset 	           & 0.0 & --- & Mean & 0.0 & -2.5 & 0.0\\
		Zero mode variance & 1e-3 & 1e-6 & Std & 2.0 & 0.5 & 3.0  \\
		Fluctuations       & 0.2 & 0.1   \\
		Flexibility        & 0.5 & 0.2   \\
		Asperity           & None & None   \\
		Average slope      & -3.0 & 1.0   \\
		\Xhline{2\arrayrulewidth}
	\end{tabular}\label{table:model_parameters_cal_OJ287}
\end{table}


\section{D-term cross correlation plots}

\begin{figure}
    \centering
    \includegraphics[width=7cm]{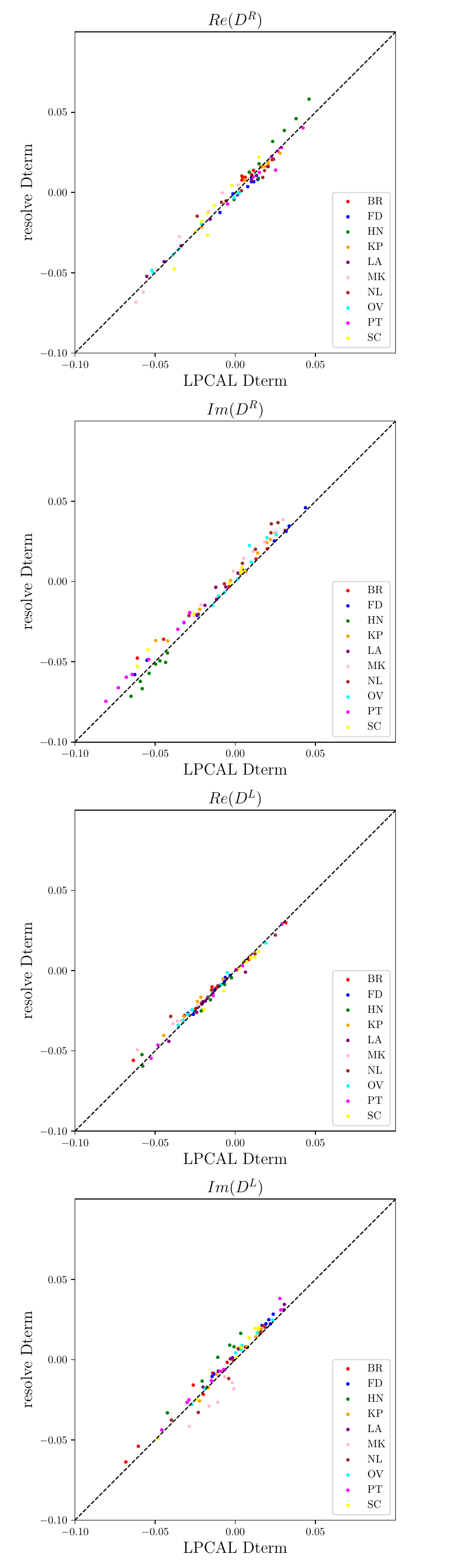}
    \caption{Cross correlation plots between 3C273 \texttt{resolve} posterior mean D-terms and \texttt{LPCAL} D-terms.}\label{3C273_15GHz_Dterm_cross_correlation}
\end{figure}

\begin{figure}
    \centering
    \includegraphics[width=7cm]{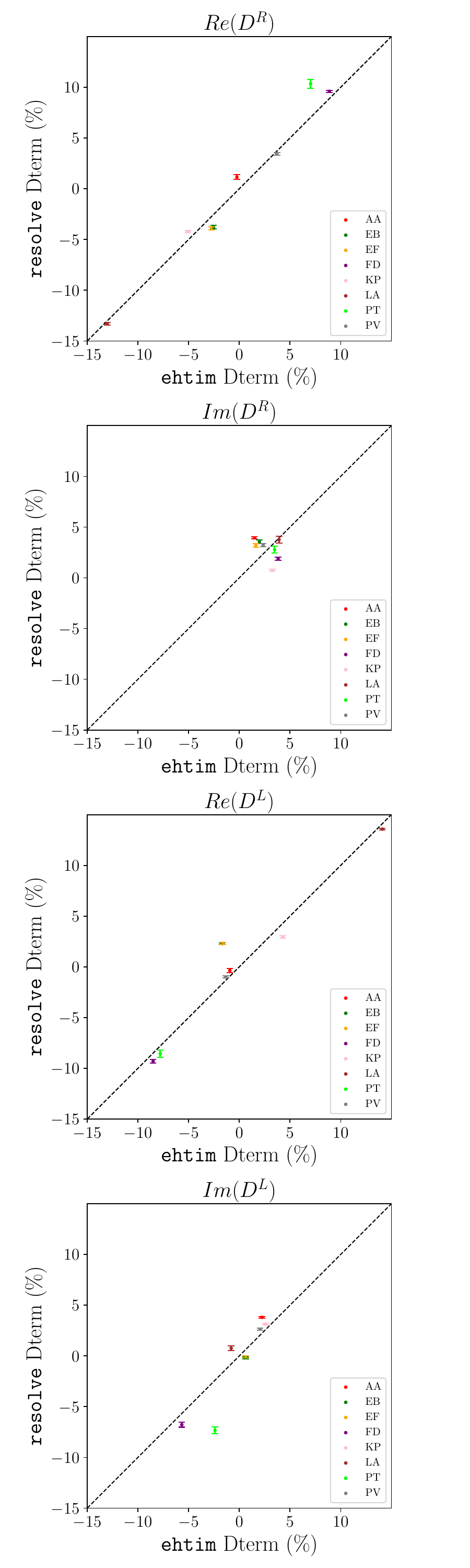}
    \caption{Cross correlation plots between OJ287 GMVA+ALMA D-term posterior means using \texttt{resolve} and D-term solutions using \texttt{ehtim}.}\label{OJ287_Dterm_cross_correlation}
\end{figure}


\section{EVPA posterior standard deviation plots}

\begin{figure}
    \centering
    \includegraphics[width=9cm]{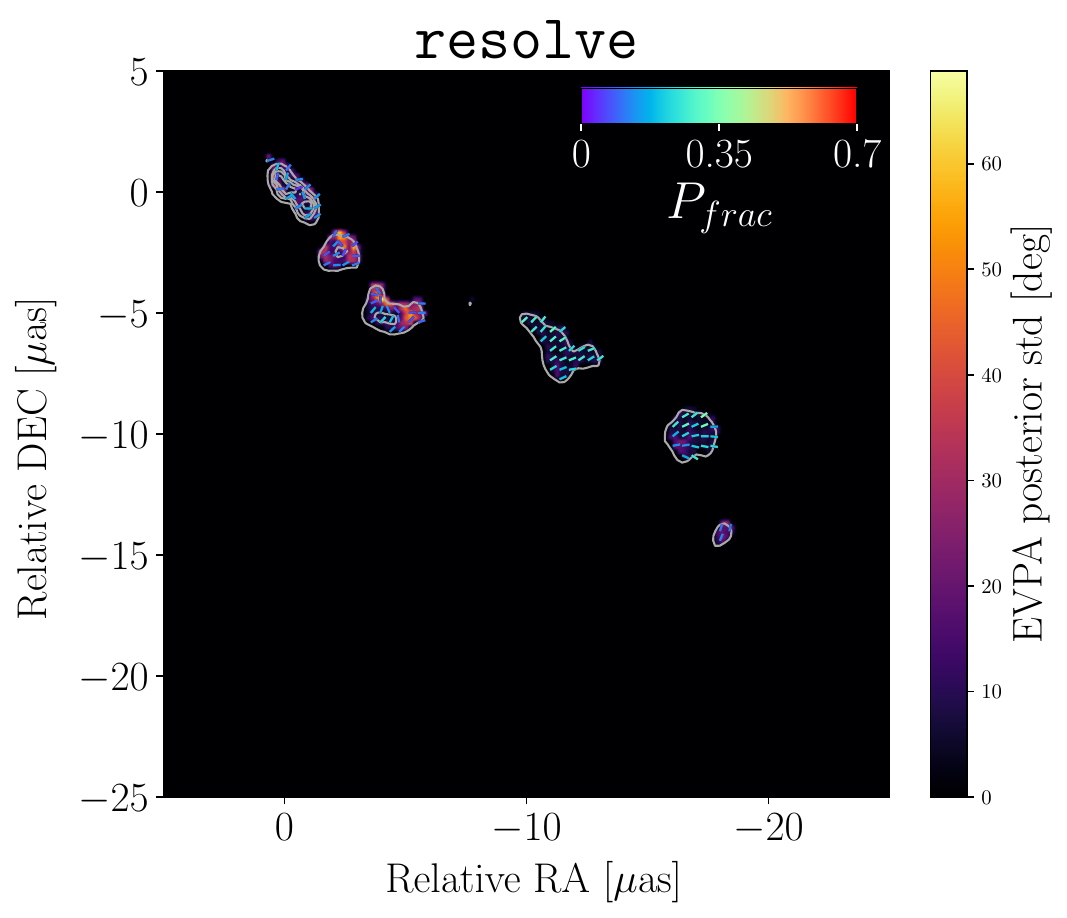}
    \caption{EVPA posterior standard deviation map of the 3C273 observation at 15GHz using \texttt{resolve}. The contours representing the total intensity \texttt{resolve} posterior mean image increase by a factor of 4, starting from 0.3$\%$ of the peak \texttt{resolve} total intensity.}\label{Fig:3C273_EVPA_std}
\end{figure}

\begin{figure}
    \centering
    \includegraphics[width=9cm]{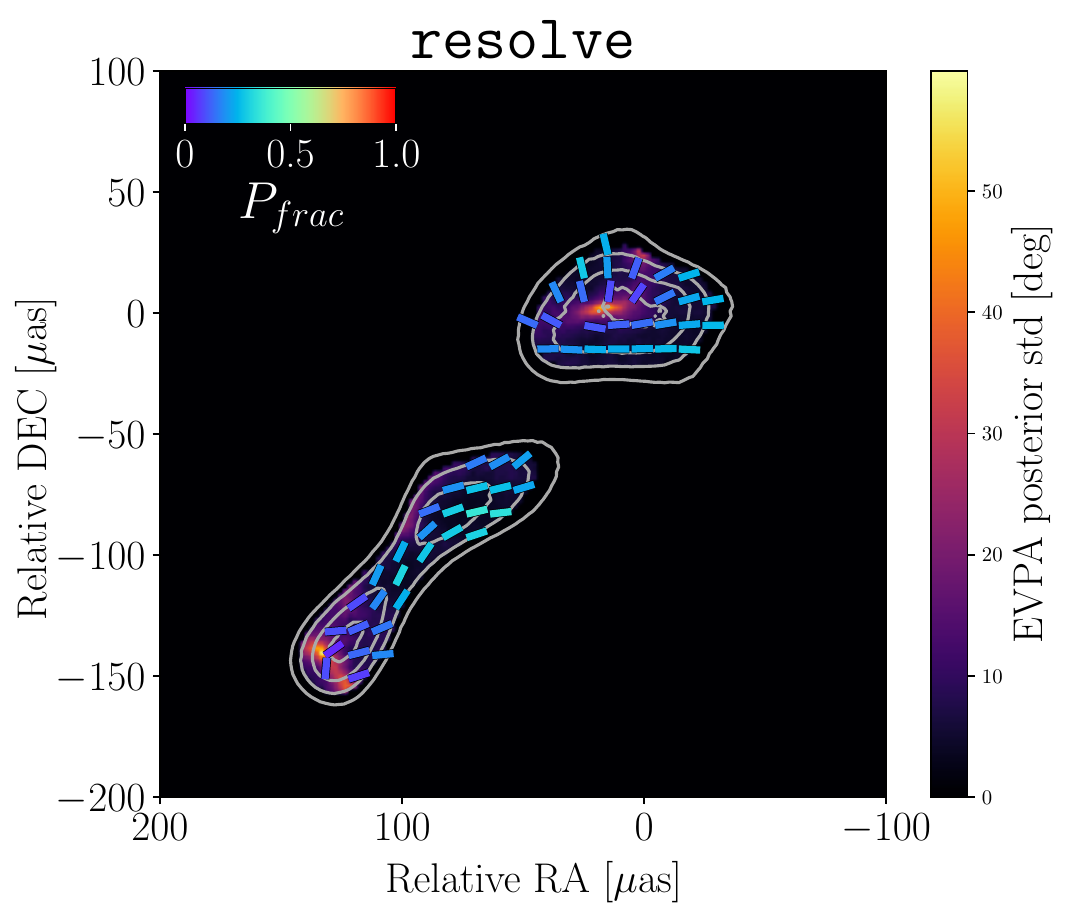}
    \caption{EVPA posterior standard deviation map of the OJ287 GMVA+ALMA observation at 86GHz using \texttt{resolve}. The contours representing the total intensity \texttt{resolve} posterior mean image increase by a factor of 2, starting from 10$\%$ of the peak \texttt{resolve} posterior mean total intensity.}\label{Fig:OJ287_EVPA_std}
\end{figure}

%
%

\end{document}